\documentclass[usenatbib]{mn2e}
\usepackage{times}
\usepackage{lscape}
\usepackage{graphicx}
\usepackage{float}
\usepackage{morefloats}
\usepackage{subfigure}
\usepackage{url}
\usepackage{rotating}
\usepackage{amssymb}

\author[S.\ Alaghband-Zadeh et al.]
{S.\, Alaghband-Zadeh,$^1$ Manda Banerji,$^{1,2}$ Paul C. Hewett,$^1$ Richard G. McMahon$^{1,2}$\\
$^1${Institute of Astronomy, University of Cambridge, Madingley Road, Cambridge, CB3 0HA, UK}\\
$^2${Kavli Institute for Cosmology, University of Cambridge, Madingley Road, Cambridge, CB3 0HA, UK}\\
}

\date{\today}
\begin{document}

\title{Heavily Reddened z$\sim$2 Type 1 Quasars II: H$\alpha$ Star Formation Constraints from SINFONI IFU Observations}
\maketitle
\begin{abstract}
We use near infrared integral field unit (IFU) spectroscopy to search for H$\alpha$ emission associated with star formation in a sample of 28 heavily reddened ($E(B-V)\simeq0.5-1.9$), hyperluminous ($\rm log(L_{bol}/ergs^{-1})\simeq47-48$) broad-line quasars at $z\simeq1.4$-$2.7$.  Sixteen of the 28 quasars show evidence for star formation with an average extinction-corrected star formation rate (SFR) of $320\pm70$\,M$_\odot$yr$^{-1}$. A stacked spectrum of the detections shows weak [NII], consistent with star formation as the origin of the narrow H$\alpha$ emission. The star-forming regions are spatially unresolved in 11 of the 16 detections and constrained to lie within $\sim$6\,kpc of the quasar emission. In the five resolved detections we find the star-forming regions are extended on scales of $\sim$8\,kpc around the quasar emission. The prevalence of high SFRs is consistent with the identification of the heavily reddened quasar population as representing a transitional phase from apparent `starburst galaxies' to optically-luminous quasars. Upper limits are determined for 10 quasars in which star formation is undetected. In two of the quasars the SFR is constrained to be relatively modest, $<$50\,M$_\odot$yr$^{-1}$, but significantly higher levels of star formation could be present in the other eight quasars. The combination of the 16 strong star formation detections and the eight high SFR limits means that high levels of star formation may be present in the majority of the sample. Higher spatial resolution data, of multiple emission lines, will allow us to better understand the interplay between star formation and  Active Galactic Nucleus (AGN) activity in these transitioning quasars.
\end{abstract}

\begin{keywords}
galaxies: active - quasars: emission lines - quasars: general 
\end{keywords}

\section{Introduction}
\label{sec:intro}
The discovery of the correlation between the bulge mass and the black-hole mass of nearby galaxies  \citep{Magorrian98, Tremaine02} lead to the prediction that the central black hole of a galaxy plays a vital role in determining the final stellar mass of the galactic bulge. Understanding this link in galaxies at $z\sim2-3$, when the majority of galaxy assembly and black-hole accretion is occurring, is therefore important to understand the formation and evolution of massive galaxies. 

In the evolutionary picture first postulated by \cite{Sanders88} infra-red luminous galaxies and quasars are different observational manifestations of the same phenomena. In the sequence illustrated by \cite{Hopkins08}, massive galaxies are thought to form hierarchically through major mergers \citep[e.g.][]{Narayanan10}. The merger induces a luminous starburst, enshrouding the galaxy in dust such that the starburst appears bright at far infra-red wavelengths. This emission is redshifted to submillimetre wavelengths for systems at $z\sim$2 and therefore the population of dusty galaxies are termed Submillimetre Galaxies (SMGs; e.g. \citealt{Blain02}, \citealt{Smail04}, \citealt{Swinbank04}, \citealt{Chapman05}). The star formation triggers black-hole accretion and as the dust and gas clear from the decaying starburst, the central nuclear region is revealed as an unobscured quasar. Feedback from the central Active Galactic Nucleus (AGN) eventually shuts off star formation within the quasar host and then the galaxy evolves passively into a red and dead elliptical \citep[e.g.][]{Springel05c,Springel05b,Croton06,Hopkins06}. 

The evolutionary scenario can be tested by observing the transition phase from massive starburst to unobscured quasar, during which the quasar will be enshrouded in dust and will therefore appear as a reddened quasar. This \textit{blowout} phase \citep{DiMatteo05} is expected to be short-lived based on estimates of the volume density of far infra-red bright quasars at $2<z<3$ ($\sim$1\,Myr estimated by \citealt{Simpson12}) and therefore large volumes must be observed to detect these systems.  

\cite{Banerji12,Banerji13} and \cite{Banerji15} (hereafter Paper I) present the results of searching for dust-reddened Type 1 broad-line quasars using the UKIRT Infrared Deep Sky Survey (UKIDSS) Large Area Survey (LAS; \citealt{Lawrence07}), the Visible and Infrared Survey Telescope for Astronomy (VISTA) Hemisphere Survey (VHS; \citealt{McMahon13}) and the Wide Infrared Survey Explorer (WISE; \citealt{Wright10}) All-Sky Survey. These studies provide a well defined sample of luminous and very massive dust-reddened quasars at $z\sim2-3$. In Paper I we find evidence that the number of reddened quasars exceeds that of the unobscured quasars at very high luminosities such that they are the dominant population of high luminosity quasars at  $z\sim2-3$, the main epoch of galaxy formation. We suggest that they represent a short phase in the evolution of massive galaxies, corresponding to high intrinsic luminosities and obscurations. 

In Paper I we utilise near infra-red Integral Field Unit (IFU) observations of the reddened quasar sample, performed using VLT-SINFONI (Spectrograph for INtegral Field Observations in the Near Infrared), by considering a single integrated spectrum per object. In this paper we now utilise the spatially resolved IFU observations to explore the star forming properties of the quasar host galaxies. Integral Field Spectroscopy provides a spectrum in each spatial pixel (spaxel) of a two dimensional field of view and the narrow H$\alpha$ emission lines in each spaxel allow the ionized gas distribution to be determined.

IFU observations of high redshift systems have been used to study the morphological and kinematic properties of star forming galaxies by tracing the narrow H$\alpha$ emission associated with star formation \citep[e.g][]{Forster-Schreiber09}. H$\alpha$ IFU observations of SMGs have found that these systems are made of multiple components, with dynamically disturbed gas motions, providing evidence that these intense starbursts are merger induced \citep{Swinbank05,Swinbank06,Alaghband-Zadeh12, Menendez-Delmestre13}. IFU observations of the [OIII]$\lambda\lambda$4959, 5007 emission-line doublet have been used to explore the dynamics of galaxies hosting AGN, finding evidence for outflows of gas \citep[e.g.][]{Alexander10,Harrison12,Harrison14}.  In the reddened quasar sample, the combination of a powerful accreting black-hole and large amounts of dust is predicted to drive strong outflows due to the effects of radiation pressure on dust \citep{Fabian12}. It is possible that outflows may both enhance or inhibit the host galaxy star formation  \citep{Silk98,Silk05,Zubovas13,Farrah12,Zinn13,Ishibashi13,Barai14}. Star formation has been found to be suppressed in the outflow regions of z$\sim$2 quasars \citep{Cano-Diaz12,Cresci14} demonstrating how AGN feedback can be probed using resolved IFU studies. 

A challenge of studying the star formation properties of quasar host galaxies is the need to separate the quasar emission (e.g. broad H$\alpha$ line) from the host galaxy emission (e.g. narrow H$\alpha$ line). Combining adaptive optics with integral field spectroscopy has proved to be effective \citep{Inskip11}. \cite{Vayner14} demonstrate that this method enables the broad H$\alpha$ emission from the quasar to be disentangled from the narrow emission associated with the host galaxy in $z\sim2$ quasars. Two quasars are found to have star formation offset from the quasar nuclei and star formation limits are constrained for the rest of the sample. In the nearby Universe, IFU observations have been used to decompose the narrow and broad H$\alpha$ emission, from star formation  and the AGN respectively, in a sample of low redshift Type 1 quasars \citep{Husemann14} resulting in the finding that over half the sample exhibit star formation on kiloparsec scales.

In this paper we explore the star formation properties of our z$\sim$2 reddened quasar sample, making use of the high resolution VLT-SINFONI observations. In Section \ref{sec:obs} we summarise the observations and reduction of the data, in Section \ref{sec:analysis} we describe the analysis techniques used to determine the star formation rates (SFRs) and place constraints on the locations of the star forming regions. In Section \ref{sec:discussion} we explore the implications of our results. All calculations assume a flat, $\Lambda$CDM cosmology with $\Omega_\Lambda=0.7$ and $H_0=71$\,kms$^{-1}$Mpc$^{-1}$ in which 0.7\,arcsec corresponds to a physical scale of $\sim$6\,kpc at $z=2.3$ (the median seeing and redshift of our observations and targets respectively).

\section{Observations and Reduction} 
\label{sec:obs}
We observed 23 reddened quasars with the near infra-red IFU VLT-SINFONI in 2013 and five with VLT-SINFONI in 2009. Details of the object selection and the observations are given in Paper I and \cite{Banerji12} respectively. Our observations targeted the redshifted H$\alpha$ emission line which lies in the $H$ and $K$ bands for our sample of sources with redshifts of $1.4<z<2.7$. The 2009 observations used the SINFONI $H$+$K$ filter (R=1500 covering a range of 1.45$-$2.45\,$\mu$m) and the 2013 observations used the $K$ filter only (R=4000 covering 1.9$-$2.5\,$\mu$m).  For both SINFONI programmes we made use of the largest field of view (8$\times$8\,arcsec) which gives a pixel scale of 0.250$\times$0.125\,arcsec which is resampled to 0.125$\times$0.125\,arcsec in the final combined cubes. Each source was moved around four quadrants of the IFU (at offsets of $\pm$1.5\,arcsec from the centre of the IFU) to enable simultaneous sky subtraction.  The resulting area with 100\% exposure was therefore 5$\times$5\,arcsec which corresponds to $\sim$40$\times$40\,kpc at $z=2.3$ (the median redshift of our sample).

The data were reduced using the standard SINFONI ESOREX pipeline which includes extraction, sky subtraction, wavelength calibration, and flat-fielding.  We spatially aligned and co-added the cubes from each observation block using an average with a 10$\sigma$ clipping threshold. We also applied a $3\times3$-pixel median filter to the co-added spatial images at each wavelength, replacing individual pixels with values $\rm |(I_{pixel}-I_{median})|$/(1.48$\times$MAD)$>$7.0 with the median value. The MAD (median absolute deviation) was calculated from the distribution of $\rm |(I_{pixel}-I_{median})|$ values for all pixels in the image.  Care was taken to ensure that the replacement threshold did not affect pixels associated with the centroids of the targets. The result of the filtering were data cubes with a vastly reduced number of spurious, single-pixel, non-Gaussian artifacts and noise. We carried out relative flux calibration by observing bright standard stars throughout the nights on which the observing blocks were performed. We were then able to perform the absolute flux calibration by comparing the integrated $K$-band fluxes of the sources to those measured from broad band VHS/UKIDSS $K$-band observations of the targets.

In Table \ref{tab:prop} we list the object identifications, redshifts and best-fit $E(B-V)$s for the targets. The error in the $E(B-V)$ values is $\sim$0.1\,mag. The $E(B-V)$ values are derived using SED fits to the available broadband photometry that probe rest-frame optical wavelengths where the choice of extinction curve makes little difference to the inferred $E(B-V)$ values. Further details of these derived properties are given in Paper I and \cite{Banerji12}.  Table \ref{tab:prop} also includes  the exposure times and the seeing of the SINFONI observations. The seeing values are derived by integrating the three dimensional IFU observations of the (unresolved) quasar emission, collapsing along the wavelength axis, to create maps of the integrated flux from which we can measure the spatial Full Width at Half Maximum (FWHM) of the Point Spread Function (PSF).  We measure the velocity resolution of the observations by determining the width of the sky lines close to the redshifted H$\alpha$ line (i.e. the velocity centre of the observations). For the 2009 SINFONI observations the resolution is $\sigma\sim$90\,kms$^{-1}$ and for the 2013 SINFONI observations the resolution is $\sigma\sim$50\,kms$^{-1}$ (Paper I).

\begin{table*}
\centering
\begin{tabular}{|l|c|c|c|c|c|c|c|c|c|c|c|}
\hline\hline
ID & RA & Dec & z &  $E(B-V)$ & Exposure & PSF &  Observation & H$\alpha$ Flux \\
&  &  &  &   & time &  FWHM & date & limit\\
\hline
 &  &  &  &   & (s) & (arcsec)  & & ($\rm \times 10^{-17}ergs^{-1}cm^{-2}$)\\ 
\hline\hline
\multicolumn{9}{|c|}{VLT-SINFONI 2013 Sample (R=4000)}\\
ULASJ0123+1525  &  20.8022 & +15.4230 &  2.629    &      1.3   & 1200   & 0.64   & 2013-07-29 & 1.7 \\
VHSJ1556-0835   & 239.1571 & -8.5952  &  2.188    &      0.7   & 1200   & 0.70   & 2013-07-29 & 0.9 \\
VHSJ2024-5623   & 306.1074 & -56.3898 &  2.282    &      0.6   & 1800   & 0.99   & 2013-07-28 & 1.3 \\
VHSJ2028-4631   & 307.0083 & -46.5325 &  2.464    &      0.6   & 1800   & 0.85   & 2013-07-30 & 2.4 \\
VHSJ2028-5740   & 307.2092 & -57.6681 &  2.121    &      1.2   & 800    & 0.71   & 2013-07-27 & 2.0 \\
VHSJ2048-4644   & 312.0435 & -46.7387 &  2.182    &      0.7   & 1600   & 0.58   & 2013-07-30 & 0.7 \\
VHSJ2100-5820   & 315.1403 & -58.3354 &  2.360    &      0.8   & 1600   & 0.87   & 2013-07-28 & 1.8 \\
VHSJ2101-5943   & 315.3311 & -59.7291 &  2.313    &      0.8   & 320    & 0.87   & 2013-07-28 & 3.7 \\
VHSJ2109-0026   & 317.3630 & -0.4497  &  2.344    &      0.7   & 1600   & 0.67   & 2013-07-30 & 1.0 \\
VHSJ2115-5913   & 318.8818 & -59.2188 &  2.115    &      1.0   & 800    & 0.79   & 2013-07-28 & 1.0 \\
VHSJ2130-4930   & 322.7490 & -49.5032 &  2.448    &      0.9   & 1600   & 0.86   & 2013-07-28 & 1.7 \\
VHSJ2141-4816   & 325.3530 & -48.2830 &  2.655    &      0.8   & 1600   & 0.65   & 2013-07-29 & 2.1 \\
VHSJ2143-0643   & 325.8926 & -6.7206  &  2.383    &      0.8   & 1600   & 0.72   & 2013-07-28 & 1.2 \\
VHSJ2144-0523   & 326.2394 & -5.3881  &  2.152    &      0.6   & 1600   & 0.71   & 2013-07-30 & 0.8 \\
VHSJ2212-4624   & 333.0796 & -46.4101 &  2.141    &      0.8   & 1800   & 0.70   & 2013-07-28 & 1.0 \\
VHSJ2220-5618   & 335.1398 & -56.3107 &  2.220    &      0.8   & 320    & 0.79   & 2013-07-29 & 1.8 \\
VHSJ2235-5750   & 338.9331 & -57.8372 &  2.246    &      0.6   & 1000   & 0.73   & 2013-07-29 & 1.2 \\
VHSJ2256-4800   & 344.1444 & -48.0088 &  2.250    &      0.6   & 1000   & 0.64   & 2013-07-29 & 1.0 \\
VHSJ2257-4700   & 344.2589 & -47.0157 &  2.156    &      0.7   & 1600   & 0.68   & 2013-07-28 & 0.9 \\
VHSJ2306-5447   & 346.5011 & -54.7881 &  2.372    &      0.7   & 1000   & 0.81   & 2013-07-28 & 1.2 \\
ULASJ2315+0143  & 348.9843 & +1.7307  &  2.560    &      1.1   & 400    & 0.46   & 2013-07-29 & 2.4 \\
VHSJ2332-5240   & 353.0387 & -52.6780 &  2.450    &      0.6   & 2400   & 0.49   & 2013-07-29 & 0.7 \\
VHSJ2355-0011   & 358.9394 & -0.1893  &  2.531    &      0.7   & 1600   & 0.54   & 2013-07-29 & 1.3 \\
\multicolumn{9}{|c|}{VLT-SINFONI 2009 Sample (R=1500)}\\                                       
ULASJ1002+0137  & 150.5470 & +1.6186   &  1.595    &      1.0   & 1200   & 1.10  & 2009-05-13 & 0.7 \\
ULASJ1234+0907  & 188.6147 & +9.1317   &  2.503    &      1.9   & 1200   & 0.76  & 2009-06-16 & 3.4 \\
ULASJ1455+1230  & 223.8375 & +12.5024  &  1.460    &      1.1   & 1200   & 0.74  & 2009-04-13 & 0.7 \\
ULASJ2200+0056  & 330.1036 & +0.9347   &  2.541    &      0.5   & 1200   & 0.66  & 2009-07-30 & 1.7 \\
ULASJ2224-0015  & 336.0392 & -0.2567   &  2.223    &      0.6   & 1200   & 0.56  & 2009-07-30 & 0.6 \\
\hline\hline
\end{tabular}
\caption{The positions, redshifts and reddenings for the targets. The error in the $E(B-V)$ values is $\sim$0.1\,mag. The top section details the newest SINFONI sample (Paper I) and the bottom section details the older SINFONI sample described in \protect\cite{Banerji12}. The redshifts are derived from the centroids of single Gaussian profiles fit to the broad H$\alpha$ lines, as described in Paper I.  We also list the exposure times of the observations and the size of the seeing disk (the FWHM in arcsec of the PSF).  We list the 1$\sigma$ flux limits of the observations, corresponding to the noise in PSF-aperture spatial regions offset from the central broad-line emission. These are calculated assuming a $\sigma$=100\,kms$^{-1}$ Gaussian profile, for a $\sigma$=200\,kms$^{-1}$ Gaussian profile these flux limits should be multiplied by $\sqrt 2$ to allow for the doubling in the wavelength region.}
\label{tab:prop}
\end{table*} 

\subsection{Noise Properties}
\label{sec:noise}
The exposure times for the VLT-SINFONI observations in 2013 were adjusted for each target at the telescope to ensure the broad H$\alpha$ line was detected. Exposure times therefore vary as a function of sky conditions, the $K$-band magnitudes and the strength of the broad H$\alpha$ emission of the individual sources (Table \ref{tab:prop}). The resulting range in spectrum flux sensitivity is significant (Table \ref{tab:prop}). As a consequence, star-formation rate sensitivities across the sample are very different, which should be noted when interpreting the results.

The wavelength-dependent noise for all the targets is dominated by the signal from the sky, with strong wavelength-dependent variations due to the presence of individual sky emission features. The noise at a given wavelength can be determined from the amplitude of flux-variations calculated using PSF-aperture spatial regions, $\sigma(\lambda)_{\rm PSF}$ that lie offset from the central AGN broad-line, i.e. excluding any 'signal' remaining after the broad-emission line subtraction for each object. The wavelength-dependence of the noise is essentially identical, with only the amplitude (from $\sigma_{\rm MAD}$\footnote{The noise amplitude for each object is calculated from the median absolute deviation of the noise, $\sigma_{\rm MAD}$, for the PSF-aperture regions, integrated across the $H$- and/or $K$-bands.}) varying from target to target.

Treating the two observing runs (2009 and 2013) separately, we constructed a master wavelength-dependent noise-spectrum by combining the results from the off-centre PSF-aperture regions of all the targets. The master noise-spectrum was then scaled using the value of $\sigma_{\rm MAD}$ for each target.

\section{Analysis and Results}
\label{sec:analysis}
To explore the star formation in the reddened quasar sample we must disentangle the H$\alpha$ emission, originating from the quasar broad line region, from any narrow H$\alpha$ line emission, originating from star formation in the quasar host galaxy.

The total emission profile is dominated by broad H$\alpha$ line emission from the quasar. The emission is unresolved and thus appears in the data cubes as a PSF, with a profile set by the observing conditions.  An example map of this broad emission, created by integrating a cube in the velocity axis over the the H$\alpha$ line (from H$\alpha$-12000\,kms$^{-1}$ to H$\alpha$+12000\,kms$^{-1}$) is shown in the top left panel in Fig. \ref{fig:cartoon}. For each source we set the central velocity (i.e. 0\,kms$^{-1}$) at the velocity centroid of the broad H$\alpha$ emission (see Paper I).  In the following analysis we search for evidence of narrow H$\alpha$ emission after subtracting the broad H$\alpha$ profile from the cubes.

\subsection{Assumptions}
\label{sec:assump}
We search for narrow emission with linewidths of $\sigma$=100\,kms$^{-1}$ and 200\,kms$^{-1}$, corresponding to narrow H$\alpha$ arising due to star formation in the host galaxy as opposed to originating from the Narrow Line Region (NLR) associated with the AGN. We perform further tests on the lines detected in Section \ref{sec:nii} to confirm the star formation origin. The $\sigma$=100\,kms$^{-1}$ and 200\,kms$^{-1}$ linewidths correspond to de-convolved linewidths of 40\,kms$^{-1}$ and 180\,kms$^{-1}$ for the 2009 observations and 90\,kms$^{-1}$ and 190\,kms$^{-1}$ for the 2013 observations.  Our choice of velocity widths are consistent with the properties of narrow H$\alpha$ observed in SMGs (with no evidence for AGN affecting the spectra) owing to the star formation in these galaxies \citep{Swinbank04,Alaghband-Zadeh12}.

Star formation rates corrected for extinction are calculated using the reddenings, given in Table \ref{tab:prop}, assuming a value of $R_v$=3.1. We make the assumption that the extinction is the same towards the quasar as towards the star forming regions. If the extinctions towards the star forming regions are lower, the derived extinction-corrected star formation rates will be over-estimated.

\subsection{Procedure}
\label{sec:pro}
\begin{figure*}
\includegraphics[width=0.95\textwidth]{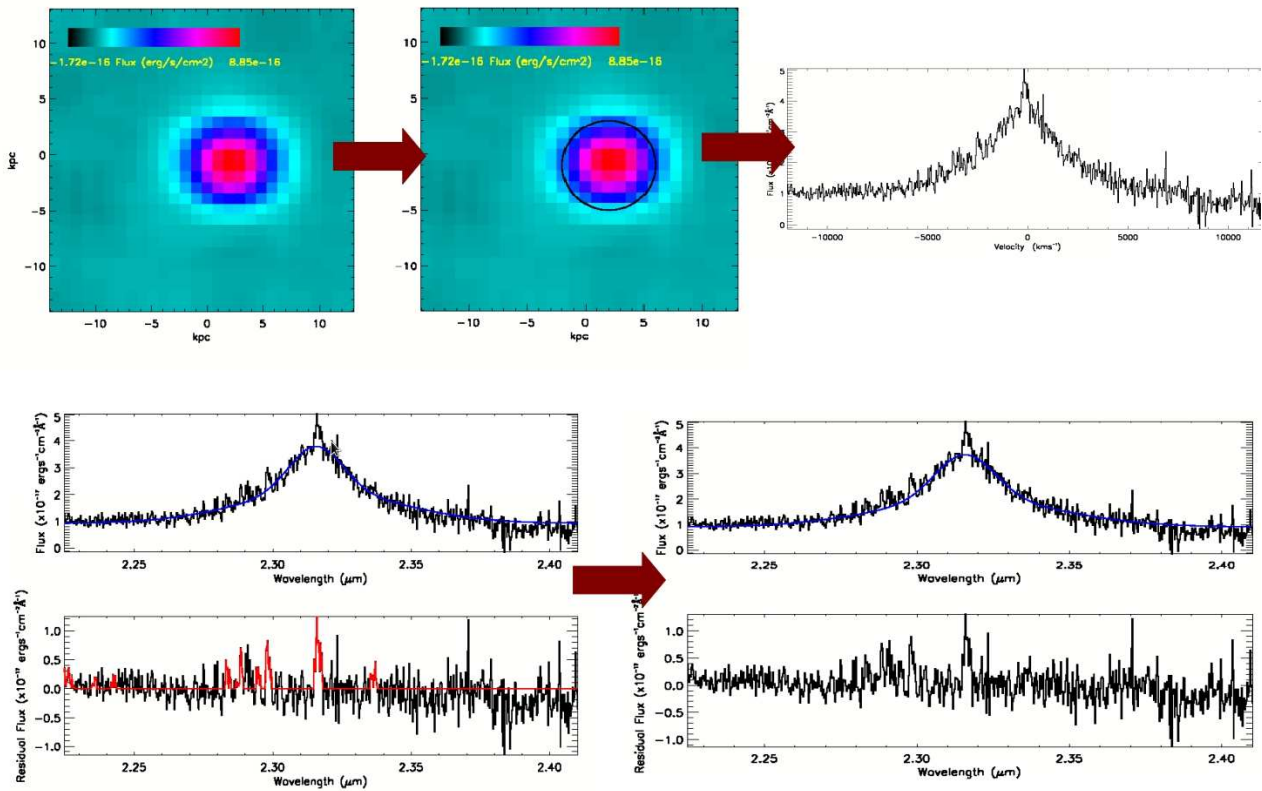}
\caption{ An example sequence (using VHSJ2355-0011) to illustrate the method used to search for narrow H$\alpha$ emission. First we integrate the cube over the broad H$\alpha$ line (from H$\alpha$-12000\,kms$^{-1}$ to H$\alpha$+12000\,kms$^{-1}$) to create an estimate of the PSF. The radius at which the PSF-signal has maximum S/N is determined and the signal from within the radius (the `PSF-aperture') is used to provide the optimal estimate of the broad line profile. We characterise the broad line profile by first fitting a single or double Gaussian profile to the spectrum (blue line). The fit requires that the Gaussian profiles must have FWHMs greater than 1000\,km$s^{-1}$. Since it is possible that the narrow component may be spatially coincident with the broad line PSF, and therefore included in this spectrum, the residuals of the initial fit are searched for regions of excess positive flux on scales of $>$\,100km$s^{-1}$ (red line). These regions are then excluded and the fit is recalculated (blue line). This final fit represents the broad line profile to be subtracted from each spaxel region. We re-bin the spaxels such that each spaxel contains the sum of the surrounding spaxels in a region the size of the PSF-aperture. We then subtract the broad profile, using a maximum likelihood technique to scale the profile, from each spaxel region and search the profile-subtracted cube for narrow Gaussian H$\alpha$ emission with $\sigma$=100\,kms$^{-1}$ and 200\,kms$^{-1}$. }
\label{fig:cartoon}
\end{figure*}

In order to search for a narrow component of H$\alpha$ we must subtract off the broad H$\alpha$ emission (Fig. \ref{fig:cartoon}). The broad H$\alpha$ emission spectrum is determined by summing the spectra of each spaxel lying within a circular aperture centred on the AGN. The radius of this circular aperture is the distance from the centre at which, for the unresolved quasar emission, the signal-to-noise ratio (S/N) within the circle is a maximum (see Fig. \ref{fig:cartoon}), therefore giving the optimal estimate of the broad line profile. We define this circular aperture as the `PSF-aperture'. We note that the `PSF-aperture' does not contain the entire flux from the quasar since the spatial profile of the quasar emission will extend beyond this region. Indeed $\simeq$32 per cent of the total flux from the unresolved emission lies outside the `PSF-aperture' and therefore an aperture correction must be applied to give the total fluxes. This correction must be applied when calculating the total fluxes of all unresolved emission, including any unresolved narrow H$\alpha$ emission we find in the following analysis.

The resulting broad H$\alpha$ emission spectra for each source are shown in Figure A2 of Paper I. We then fit a single or double broad-Gaussian profile to each spectrum. The FWHMs of the single profile or both components of the double profile are constrained to be $>$1000\,kms$^{-1}$ and $<$10000\,kms$^{-1}$ in the fit. We choose to use a double Gaussian profile only in the cases where the fit is significantly improved by the addition of the second component, as in Paper I. The FWHMs of the broad line fits range from 1000$-$9600\,kms$^{-1}$.

Any narrow H$\alpha$ emission may be coincident in velocity with the broad H$\alpha$ profile and therefore included in the overall Gaussian fits mentioned above. To identify any such emission we search the residuals from the broad Gaussian fit for regions of excess positive flux on scales of $\sim$100\,kms$^{-1}$ and exclude these spectral regions in the next fit iteration (illustrated in Fig. \ref{fig:cartoon}). This final fit represents the defined broad line profile to be subtracted from the spectra across the cube.

We re-bin the cube such that the spectrum in each spaxel is replaced with the sum of the spectra from the spaxels in the surrounding region of the size of the PSF-aperture. In this way we therefore search for narrow H$\alpha$ emission in PSF-apertures across the field of view. The broad-line profile is then subtracted from each binned spaxel in the data cube with an amplitude, $k$, determined using the maximum-likelihood technique \citep[e.g.][]{Hewett85} where

\begin{equation} 
k= \frac{\Sigma_i(S_iP_i)/\sigma_i^2}{\Sigma_i(P_i/\sigma_i)^2}
\label{eq:kscale} 
\end{equation}
 
and $S_i$ is the spectrum (with $i$ wavelength steps) in each spaxel, $P_i$ is the broad profile defined above and $\sigma_i$ is the wavelength dependent noise (Section \ref{sec:noise}).

\begin{figure}
\includegraphics[width=0.45\textwidth]{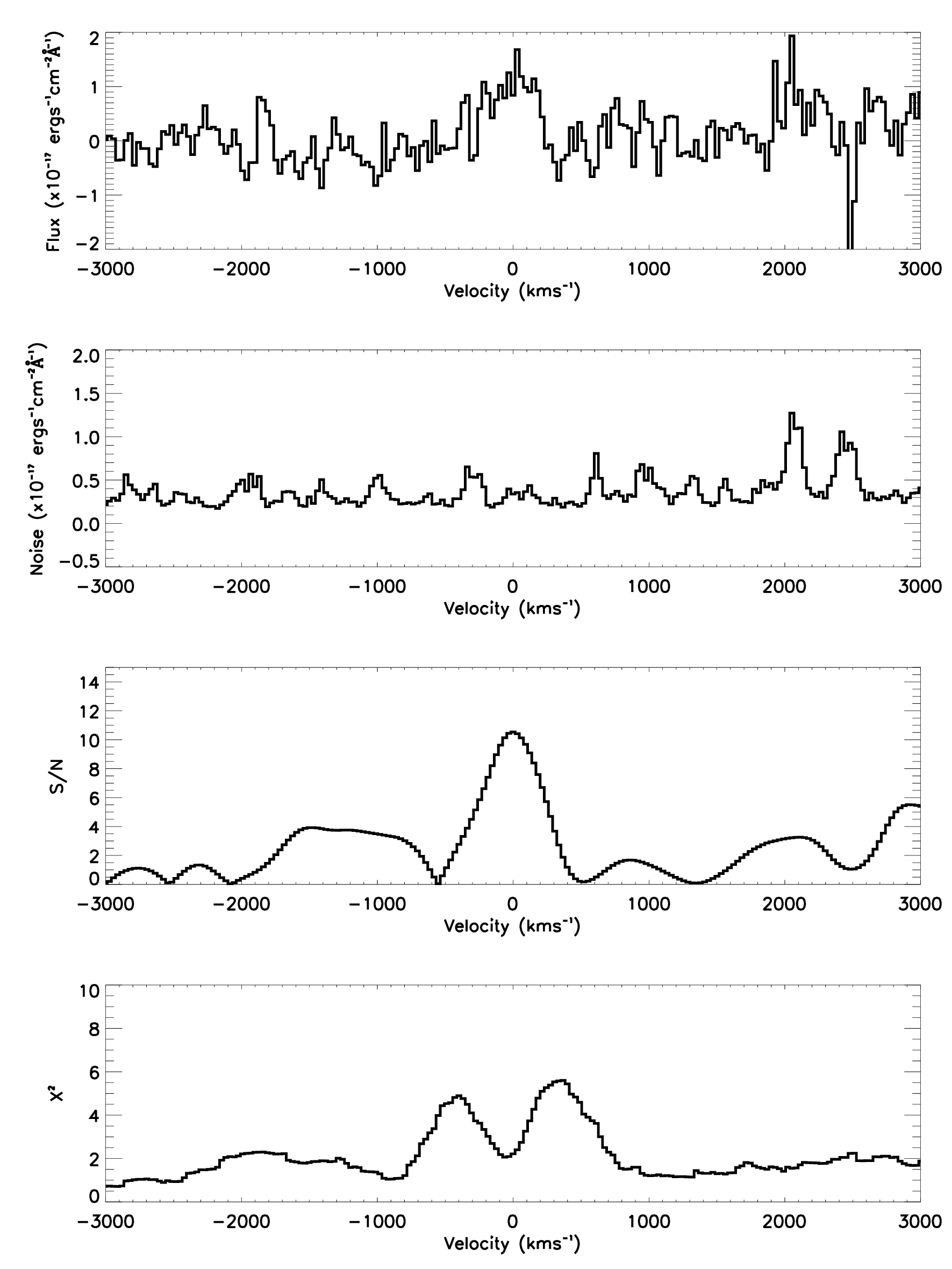}
\caption{ Top: An example of a narrow H$\alpha$ line (from VHSJ2141-4816) originating from a PSF-aperture. Second from Top: The corresponding wavelength dependent noise. 3rd from Top: The S/N at each velocity of a $\sigma$=200\,kms$^{-1}$ Gaussian, scaled to fit the spectrum at that point. Bottom: The corresponding reduced $\chi^2$ at each velocity of a $\sigma$=200\,kms$^{-1}$ Gaussian. The location of the line centroid is the velocity where the $\chi^2$ reaches a local minimum (and is below 4) and the S/N$>$5.}
\label{fig:2141_example}
\end{figure}

After subtracting the broad-line profile from each binned spaxel we search the resultant residual spectra for narrow H$\alpha$ emission, assumed to possess a Gaussian profile with widths of $\sigma$=100\,kms$^{-1}$ or 200\,kms$^{-1}$ (see Sections \ref{sec:assump} and \ref{sec:crit}). At each wavelength, the amplitude and $\chi^2$ of the two Gaussian profiles is calculated, using velocity intervals of 600\,kms$^{-1}$ or 1200\,kms$^{-1}$ (for the 100\,kms$^{-1}$ or 200\,kms$^{-1}$ search respectively). Fig. \ref{fig:2141_example} shows how the S/N and $\chi^2$ values change across an example spectrum containing a narrow H$\alpha$ line when searching for a $\sigma$=200\,kms$^{-1}$ Gaussian profile.

\subsubsection{Criteria for detections}
\label{sec:crit}
\begin{figure*}
\includegraphics[width=0.95\textwidth]{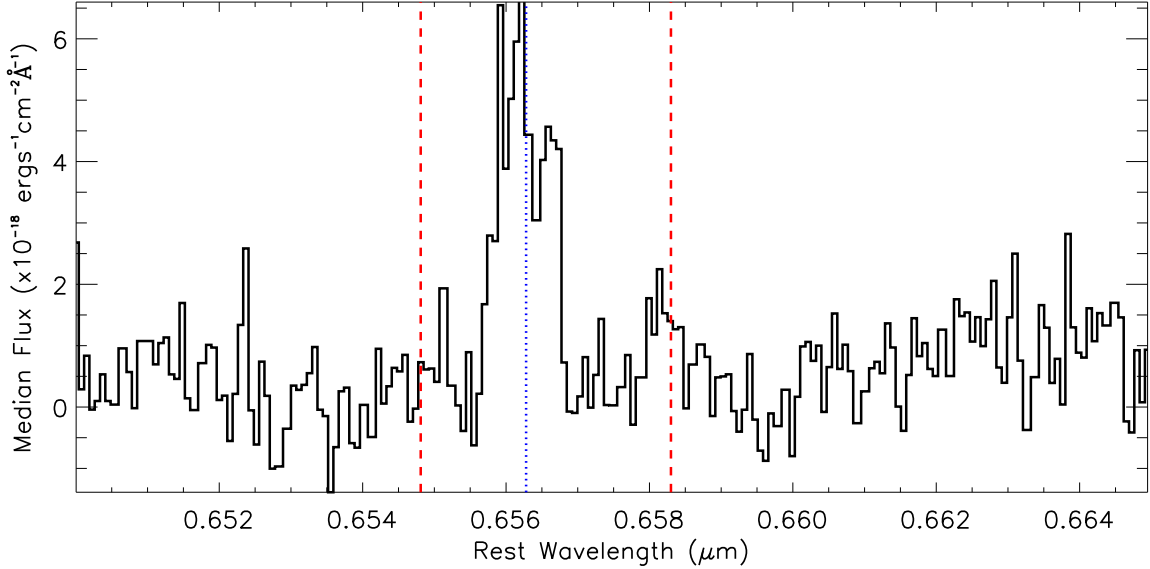}
\caption{The stacked spectra of 14 detections of narrow H$\alpha$ emission (marked by the blue line) with the position of the [NII] lines marked by the red lines. We do not include the 2 high S/N narrow H$\alpha$ detections (VHSJ1556-0835 and ULASJ2200+0056) as discussed in Section \protect\ref{sec:nii}. The ratio of [NII]/H$\alpha$ in the stack is 0.25$\pm$0.08.}
\label{fig:stack}
\end{figure*}

H$\alpha$ emission from star-formation is expected to be present at relatively low S/N. The number of potential locations in position and velocity searched is large: $\sim$80000 locations within $\pm$1000\,kms$^{-1}$ of the systemic velocity and $\pm$20\,kpc from the location of the quasar centre for each source are searched in the 2013 sample. In the 2009 sample, with the lower spectral resolution, $\sim$30000 locations are searched for each source. False-positive detections are of particular concern, therefore we choose a conservative S/N threshold of  S/N$>$5 for our narrow H$\alpha$ detections. The appropriateness of the S/N$>$5 threshold was verified from stacking the narrow H$\alpha$ detections with S/N$>$5 (Fig. \ref{fig:stack}) and finding evidence for the presence of a weak [NII]$\lambda$0.6583$\mu$m signal (see Section \ref{sec:nii}).

The Gaussian-profile template search, with the two velocity widths $\sigma$=100 and 200\,kms$^{-1}$, retains sensitivity to the presence of all emission features with widths $\sigma \lesssim$250\,kms$^{-1}$. The $\chi^2$ values do, however, increase for features that do not match the Gaussian profiles of either specific velocity-width. A relaxed $\chi^2$-threshold of $\chi^2<$4 was adopted to ensure any such features were not excluded from the search.

The area with 100\,per cent exposure in our observations is 5$\times$5\,arcsec and our emission-line search extends over an area of 4$\times$4\,arcsec centred on the quasar, which corresponds to $\sim$30$\times$30\,kpc at $z$=2.3 (the median redshift of our sample).  Since we search in overlapping PSF-apertures, real emission lines result in multiple detections in neighbouring spaxels.  In such cases we first take the location of the line detection to be the spaxel with the highest S/N. After examination of these detections, we find that the locations of all the detections are within the PSF-sized aperture region centred around the quasar emission. The narrow emission is therefore consistent with lying at the centre of the quasar and we therefore present the spectra from the central PSF-apertures as the detections in the following analysis.

\begin{figure}
\includegraphics[width=0.45\textwidth]{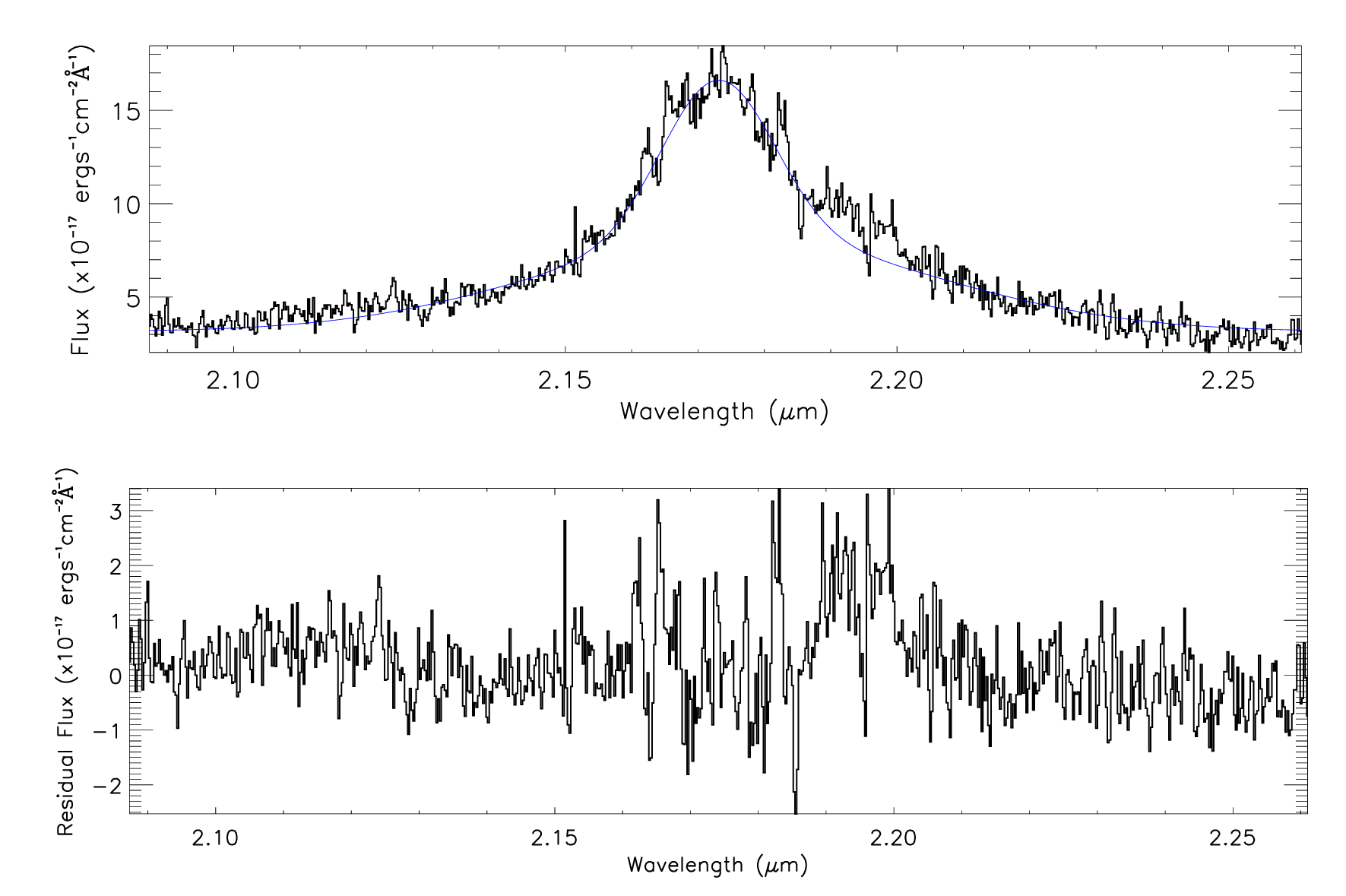}
\caption{Top: The broad H$\alpha$ profile of VHSJ2101-5943 with the double Gaussian fit overlayed (blue). Bottom: The residuals of the fit to the broad H$\alpha$ profile. }
\label{fig:2101}
\end{figure}

\begin{figure}
\includegraphics[width=0.45\textwidth]{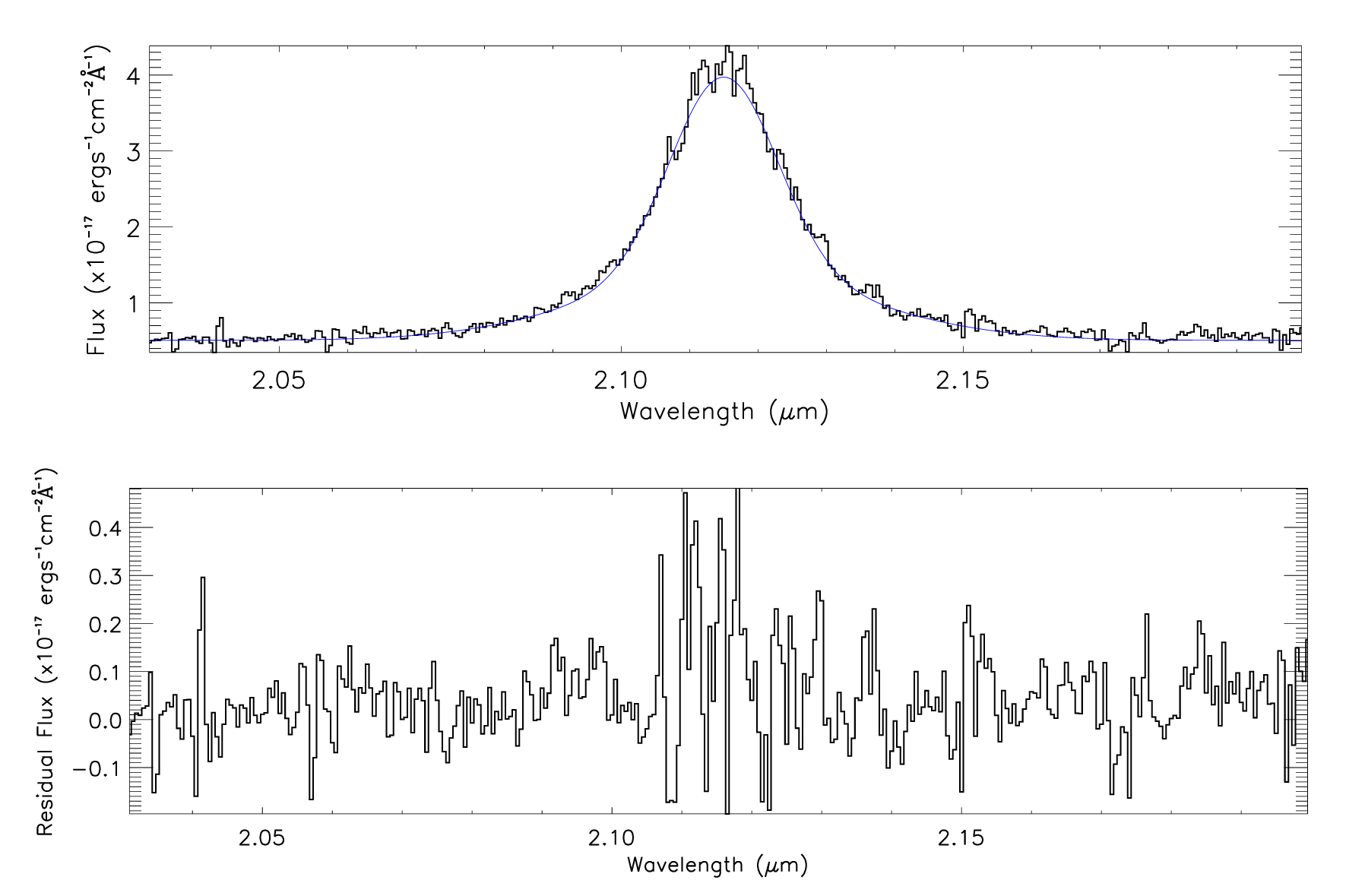}
\caption{Top: The broad H$\alpha$ profile of ULASJ2224-0015 with the double Gaussian fit overlayed (blue). Bottom: The residuals of the fit to the broad H$\alpha$ profile. }
\label{fig:2224}
\end{figure}

After running the automated search for narrow H$\alpha$ emission, with the above detection criteria, we apply an additional check of the broad H$\alpha$ profile subtraction, examining the residuals of the fit. In two of the sample (VHSJ2101-5943 and ULASJ2224-0015) potential narrow H$\alpha$ emission is found, meeting our detection criteria. However, after subtracting the broad H$\alpha$ profile we find that the resulting spectra contain multiple large residuals in the region of H$\alpha$$\pm$5000\,kms$^{-1}$, suggesting that the narrow emission features are not real. In Figs. \ref{fig:2101} and \ref{fig:2224} we show the residuals of the broad subtractions for VHSJ2101-5943 and ULASJ2224-0015 respectively. The potential narrow emission found in our automated search for VHSJ2101-5943 and ULASJ2224-0015 is shown in Figs. \ref{fig:narrow_VHSJ2101-5943} and  \ref{fig:narrow_ULASJ2224-0015} respectively, however we exclude these sources from the narrow H$\alpha$ detection sample for the remainder of the analysis. 

\subsubsection{Search for extended emission}
\label{sec:extended}
It is possible that the narrow emission we detect originates from star forming regions that are extended beyond the size of the PSF-apertures. In order to test for resolved, extended emission in the sample of narrow  H$\alpha$ detections we increase the aperture size of the binned cubes and re-run our narrow H$\alpha$ emission search procedure. In ULASJ1002+0137, VHSJ2028-4631, VHSJ2144-0523,  VHSJ2235-5750 and VHSJ2332-5240 the S/N of the resulting narrow emission is improved by increasing the aperture size and therefore we find five sources in which there is evidence for extended and resolved star forming regions lying around the quasar centres. The sizes of the larger apertures (listed in Table \ref{tab:sfr_in}) correspond to the aperture sizes at which the S/N of the detections are maxima. In the analysis that follows we use the spectra from these larger apertures.

\subsubsection{Search for [NII]}
\label{sec:nii}

Application of the search procedure described above results in the detection of narrow H$\alpha$ emission in 16 of the 28 reddened quasars. In order to test whether the narrow emission arises due to star formation in the quasar host galaxy or from the NLR of the AGN itself we search the spectra for [NII] emission. Strong [NII] would indicate that the narrow emission may originate from the NLR, whereas weak [NII] would suggest a star-formation origin. Given the velocity centroid of the narrow H$\alpha$ emission we can constrain the location of the [NII] emission in each spectrum.  We search only for the stronger [NII] $\lambda_{rest}$=0.6583$\mu$m component of the [NII] doublet given its suitability for discriminating between star-formation and AGN origin of the H$\alpha$ emission \citep[e.g.][]{Kewley06}.  In each of the 16 spectra, based on the velocity-width of the putative H$\alpha$ detection, we fit a $\sigma$=100\,kms$^{-1}$ or 200\,kms$^{-1}$, Gaussian profile at the predicted position of the [NII] line. 

Two of the sources (VHSJ1556-0835 and ULASJ2200+0056) exhibit large residuals in the spectral region of [NII] following the broad H$\alpha$ profile subtraction (shown in Figs. \ref{fig:narrow_VHSJ1556-0835} and \ref{fig:narrow_ULASJ2200+0056}). Given the high S/N detections of the narrow H$\alpha$ in these two sources we are confident that the residuals at the position of [NII] are not real [NII] features. The narrow H$\alpha$ emission in both cases is Gaussian shaped, whereas the excess flux at the position of [NII] is highly non-Gaussian, therefore not matching the narrow H$\alpha$ line features. For these two sources we are therefore unable to constrain the real [NII] emission. The origin of the excess flux may be from poor characterisation of the broad H$\alpha$ line, for example if it is not well fit by a single or double Gaussian profile, such that the broad subtraction leaves excess flux at the location of [NII]. Narrow [NII] associated with the narrow H$\alpha$ may be present but the extent of the residuals precludes any reliable measurement. In the analysis that follows we include VHSJ1556-0835 and ULASJ2200+0056 in our sample of star-formation detections, since we cannot rule out that the narrow H$\alpha$ emission arises due to star formation, however we do not include these two sources in the stack of the detections (Fig. \ref{fig:stack}) since we use the stack to constrain the low S/N [NII] emission for the sample.

The stack of the 14 remaining detections is shown in Fig. \ref{fig:stack} and we constrain the [NII]/H$\alpha$ ratio for the stack as [NII]/H$\alpha$=0.25$\pm$0.08. The low [NII]/H$\alpha$ ratio is consistent with a star-formation origin (e.g. \citealt{Kewley06} who find [NII]/H$\alpha$$<$0.7 as indicative of star formation).

For VHSJ2235-5750 we detect [NII] with a S/N$>$5 and therefore we can establish the [NII]/H$\alpha$ ratio of 0.45$\pm$0.09. Since this ratio is $<$0.7 we are confident of the star-formation the origin of the narrow H$\alpha$ emission for this source. For the 13 remaining detections we quote the limits on the [NII]/H$\alpha$ ratio in Table \ref{tab:sfr_in}. The upper limit of the [NII] flux is the 5$\sigma$ limit, from our S/N$>$5 detection criteria. In nine of the detections the upper limits to the [NII]/H$\alpha$ ratio are $<$0.7 indicating that the narrow H$\alpha$ emission is from star formation. In the remaining four detections the upper limits to the [NII]/H$\alpha$ ratio are $>$0.7 and it is possible that the narrow H$\alpha$ emission originates from the NLR, however, since only upper limits are available we retain the sources in our sample of objects possessing evidence of star formation.

\subsection{Results} 
\label{sec:results}
\begin{figure*}
\includegraphics[width=0.95\textwidth]{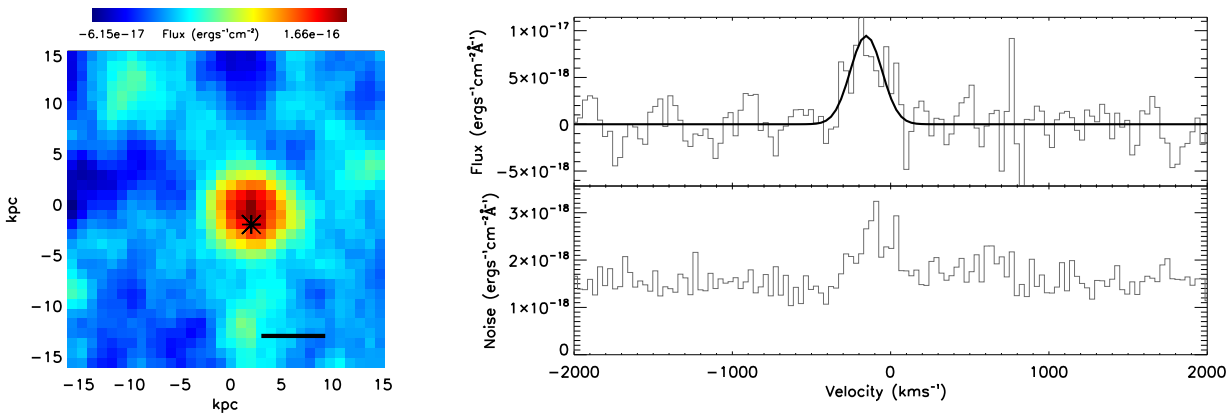} 
\caption{Left: Map of the narrow H$\alpha$ emission in VHSJ2355-0011, created by integrating the profile-subtracted cube from the centre of the narrow line $\pm$150\,kms$^{-1}$ (1.5$\sigma$). The profile-subtracted cube is made by subtracting the broad line profile (scaled using the maximum likelihood technique) from the spectrum in each binned spaxel.  The spaxels are binned such that each spaxel represents the sum of the surrounding spaxels in a PSF-aperture sized region. The black cross marks the quasar centre and corresponds to the aperture in which the total spectrum has evidence for a narrow Gaussian of width $\sigma$=100\,kms$^{-1}$. We show the FWHM of the PSF-aperture by the black line in the bottom, right-hand corner. Top Right: The total spectrum from the aperture at the quasar centre with the scaled Gaussian overlayed. Bottom Right: The noise spectrum (as detailed in Section \protect\ref{sec:noise}).}
\label{fig:VHSJ2355-0011_example}
\end{figure*}

In Figs. \ref{fig:narrow_ULASJ1002+0137} to \ref{fig:narrow_VHSJ2355-0011} we show the maps and spectra of the 16 star formation detections, along with the corresponding noise spectra. In Fig \ref{fig:VHSJ2355-0011_example} we show an example detection of the narrow emission in VHSJ2355-0011. The maps of the detections are made by integrating the broad H$\alpha$ profile-subtracted cubes from the centre of the narrow line $\pm$150\,kms$^{-1}$ or $\pm$300\,kms$^{-1}$, corresponding to 1.5$\sigma$ for the 100\,kms$^{-1}$ and 200\,kms$^{-1}$ search respectively. The profile-subtracted cubes are made by subtracting the broad line profile (scaled using the maximum likelihood technique) from the spectrum in each binned spaxel. The spaxels are binned such that each spaxel represents the sum of the surrounding spaxels in a PSF-aperture sized region.  For each detection we present the spectrum originating from the binned spaxel at the quasar centre, in which there is evidence for narrow H$\alpha$ emission.

Table \ref{tab:sfr_in} lists the extinction-corrected star-formation rates (assuming the \cite{Calzetti00} reddening law) using the relation between H$\alpha$ luminosity and star formation rate from \cite{Kennicutt12}:

\begin{equation}
\rm{log(SFR_{H\alpha}/M_{\odot} yr^{-1})}=log(L_{H\alpha}/erg s^{-1}) - 41.27
\label{eq:SFR_Ha}
\end{equation} 

In Table \ref{tab:sfr_in} we quote the star formation rates using the H$\alpha$ flux within the PSF-apertures for the unresolved sample (SFR$\rm_{aperture}$). The PSF-aperture is defined as a circle with a radius at which, for an unresolved point source, the S/N of the emission is a maximum (see Section \ref{sec:pro}) and $\simeq$32 per cent of the total flux lies outside the aperture. We therefore apply an aperture-correction to calculate the total star formation rate (SFR$\rm_{total}$).  For objects with resolved detections, the H$\alpha$ flux within the extended aperture is used to derive the star formation rate.  Sizes, in kpc, of the star forming regions are also given.  The sizes correspond to the FWHMs of the apertures for the resolved detections and to the FWHMs of the PSF-apertures (as upper limits) for the unresolved detections.

While we explore the spatial positions of the narrow H$\alpha$ detections we do not analyse the velocity offsets of the narrow H$\alpha$ emission from the systemic velocities since the centroids of the broad H$\alpha$ emission are themselves not tightly constrained.

\begin{table*}
\centering
\begin{tabular}{|l|c|c|c|c|c|c|c|c|c|c|c|}
\hline\hline
ID & Flux$\rm_{aperture}$ & SFR$\rm_{aperture}$  & SFR$\rm_{aperture}$  & SFR$\rm_{total}$   & Size & [NII]/H$\alpha$\\
 &  & (uncorrected for extinction) & (extinction corrected) & (extinction corrected)  &  & \\
\hline
& ($\rm \times 10^{-17}ergs^{-1}cm^{-2}$) & ($\rm M_{\odot}$yr$^{-1}$) & ($\rm M_{\odot}$yr$^{-1}$) & ($\rm M_{\odot}$yr$^{-1}$) & (kpc) & \\
\hline\hline
ULASJ0123+1525& $-$ & $-$  & $-$& $<$1100  & $<$5.1 & $-$  \\
ULASJ1002+0137& 5$\pm$1 & 4.5$\pm$0.9 & 47$\pm$9 & 47$\pm$9 & 11$\pm$6 & $<$1.1\\
ULASJ1234+0907& $-$  & $-$  & $-$& $<$4600   & $<$6.1 & $-$  \\
ULASJ1455+1230& $-$  & $-$  & $-$& $<$190    & $<$6.3  & $-$  \\
VHSJ1556-0835 & 11.9$\pm$0.8 & 23$\pm$2 & 119$\pm$8 & 160$\pm$10 &  $<$5.8 & $-$\\
VHSJ2024-5623 & $-$  & $-$  & $-$& $<$110    & $<$8.1  & $-$  \\
VHSJ2028-4631 & 31$\pm$4  & 80$\pm$10 & 320$\pm$50 &   320$\pm$50  & 11$\pm$1 & $<$0.7 \\
VHSJ2028-5740 & 20$\pm$2  & 36$\pm$3 &  600$\pm$50 &   820$\pm$70 & $<$5.9 & $<$0.7 \\
VHSJ2048-4644 & 7.8$\pm$0.6 & 15$\pm$1 & 78$\pm$6 &  90$\pm$7  & $<$4.8 & $<$0.28 \\
VHSJ2100-5820 & 34$\pm$3 & 78$\pm$6 & 510$\pm$40 &  670$\pm$50 & $<$7.1 & $<$0.29 \\
VHSJ2101-5943 & $-$ & $-$ &  $-$ &   $<$190  & $<$7.1 & $-$ \\
VHSJ2109-0026 & $-$  & $-$  & $-$& $<$120    & $<$5.5  & $-$  \\
VHSJ2115-5913 & 7.3$\pm$0.8 & 13$\pm$1 & 130$\pm$10 &   200$\pm$20  & $<$6.6 & $<$1.1 \\
VHSJ2130-4930 & 17$\pm$1  & 42$\pm$3 & 350$\pm$30  & 450$\pm$30  & $<$7.0 & $<$0.29 \\
VHSJ2141-4816 & 37$\pm$4 & 120$\pm$10 &  750$\pm$70 &   940$\pm$90  & $<$5.2 & $<$0.6 \\
VHSJ2143-0643 &  5$\pm$1 & 12$\pm$2  & 80$\pm$20 & 110$\pm$20    & $<$5.9  & $<$1.0  \\
VHSJ2144-0523 & 18$\pm$2 & 33$\pm$4 &    130$\pm$10 & 130$\pm$10  & 7$\pm$4 & $<$0.41 \\
ULASJ2200+0056& 43$\pm$2 & 121$\pm$5 &   390$\pm$20 &  490$\pm$20  & $<$5.3 & $-$ \\
VHSJ2212-4624 & 4.1$\pm$0.8  & 8$\pm$1   & 49$\pm$9 & 70$\pm$10   & $<$5.8  & $<$0.9  \\
VHSJ2220-5618 & $-$  & $-$    & $-$ &$<$160   & $<$6.5  & $-$  \\
ULASJ2224-0015& $-$  & $-$    & $-$ &$<$60   & $<$4.6  & $-$  \\
VHSJ2235-5750 & 18$\pm$2 & 37$\pm$4 &    150$\pm$20 & 150$\pm$20  & 8$\pm$2 & 0.45$\pm$0.09 \\
VHSJ2256-4800 & $-$  & $-$    & $-$& $<$40   & $<$5.3  & $-$  \\
VHSJ2257-4700 & $-$  & $-$    & $-$ &$<$24   & $<$5.6  & $-$  \\
VHSJ2306-5447 & $-$  & $-$    &  $-$&$<$150   & $<$6.6  & $-$  \\
ULASJ2315+0143& $-$ & $-$ &   $-$ & $<$180 & $<$3.7 & $-$ \\
VHSJ2332-5240 & 12$\pm$1 & 31$\pm$3 &    130$\pm$10 & 130$\pm$10  & 5$\pm$2 & $<$0.49 \\
VHSJ2355-0011 & 18$\pm$2 & 51$\pm$6 &    260$\pm$30 & 290$\pm$30  & $<$4.4 & $<$0.34 \\
\hline\hline
\end{tabular}
\caption{The fluxes and star formation rates of the narrow H$\alpha$ emission. We quote both the extinction corrected star formation rates and the rates uncorrected for extinction. For the unresolved detections we also scale the extinction corrected SFRs within the PSF-apertures to give the total aperture-corrected SFRs. For the cases where we do not see any evidence for narrow H$\alpha$ emission we quote the input SFR required to be added to the cubes (in the form of a narrow Gaussian component) for the narrow component to be detected in our procedure (i.e. with S/N$>$5) as detailed in Section \ref{sec:sim}. For the unresolved detections we quote the FWHM of the PSF in kpc as the limit on the size of the star forming regions around the quasar centres.The FWHM of the PSF is listed in arcsec in Table \protect\ref{tab:prop}. For the resolved emission we quote the FWHM of the aperture size that maximises the S/N of the detections. We quote the upper limits to the [NII]/H$\alpha$ ratio for the sources with narrow H$\alpha$ detections as detailed in Section \ref{sec:nii}. The upper limit to the [NII] flux is the 5$\sigma$ upper limit such that we match the S/N$>$5 detection criteria.}
\label{tab:sfr_in}
\end{table*}

\begin{figure*}
\centering
\includegraphics[scale=0.6,angle=0]{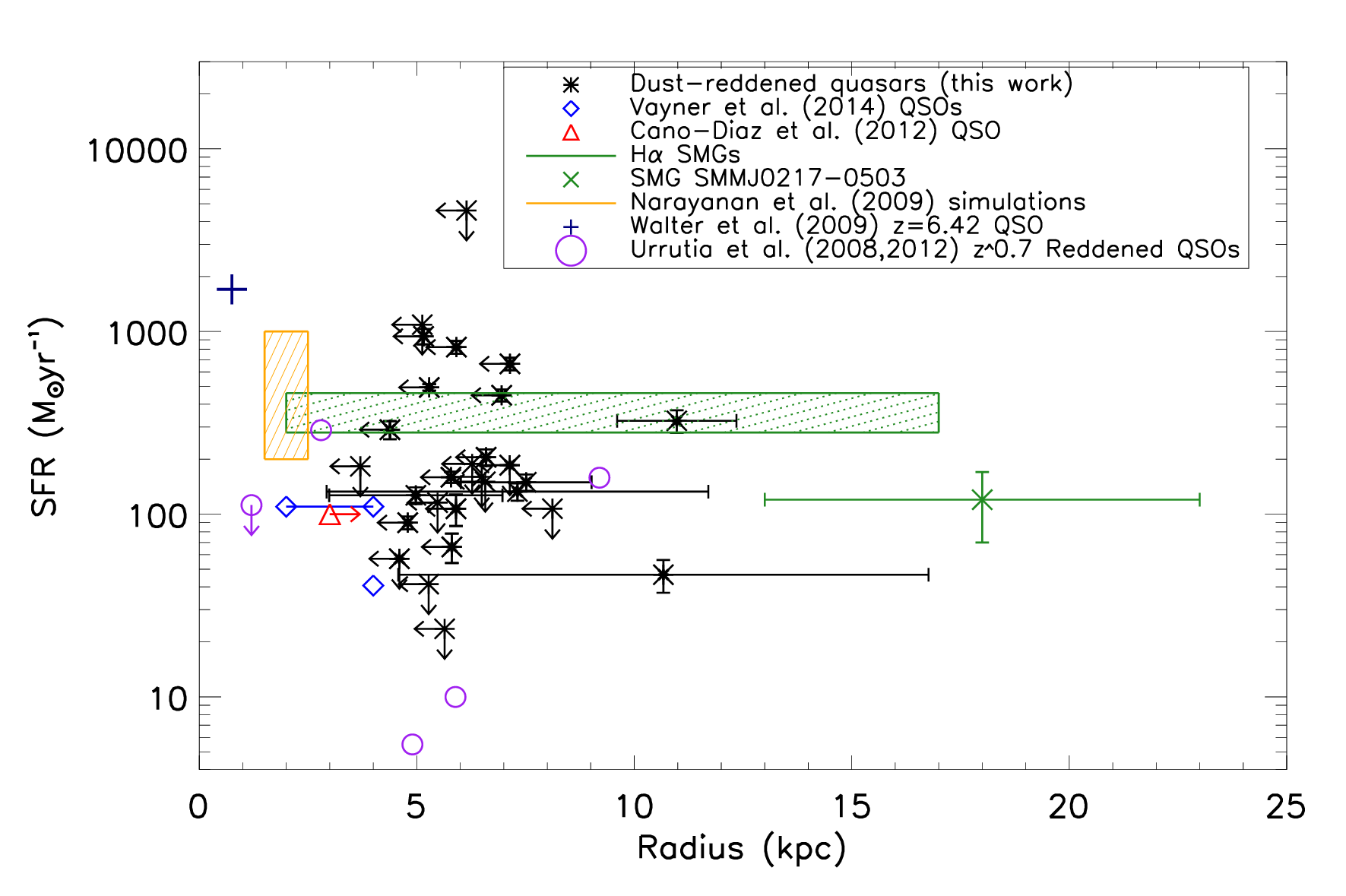}
\caption{The total, i.e. aperture-corrected in the case of unresolved emission, star-formation rates derived from the narrow H$\alpha$ emission as a function of the radius within which the narrow H$\alpha$ emission is constrained to originate. The radii correspond to the FWHM of the PSF of the broad H$\alpha$ emission for the unresolved emission and the aperture size that maximises the S/N for the resolved emission.  For sources without H$\alpha$ emission detections we plot an upper limit to the star-formation rate calculated as described in Section \ref{sec:sim}. We plot literature detections of star formation around other high-z quasars (\protect\citealt{Vayner14,Cano-Diaz12,Walter09}), $z\sim$0.7 reddened quasars (\protect\citealt{Urrutia08,Urrutia12}) and also the predictions from hydrodynamical simulations of mergers (\protect\citealt{Narayanan09}. We also show the range of extents of the star formation in SMGs from H$\alpha$ observations (\protect\citealt{Swinbank06,Alaghband-Zadeh12,Menendez-Delmestre13}) and the average H$\alpha$ star formation rate of SMGs from (\protect\citealt{Alaghband-Zadeh12}). We separately plot the offset and star formation rate of the SMG SMMJ0217-0503 which represents a SMG and AGN system, clearly offset in the H$\alpha$ observations (\protect\citealt{Alaghband-Zadeh12}). The scale of star formation observed in the majority of our reddened quasar sample is consistent with the observations of other high-z quasars and the local reddened quasars.}
\label{fig:sfr_r}
\centering
\end{figure*}

In Fig. \ref{fig:sfr_r} we show the star formation rates against the star forming region sizes. The average star formation rate is 320$\pm$70\,$\rm M_{\odot}$yr$^{-1}$ for the 16 quasars in which we detect narrow H$\alpha$ emission.  We find extended, resolved emission in five sources with star forming regions of 8$\pm$1\,kpc around the quasar centre. The emission is unresolved in the remainder of the sample, typically constrained to lie within $\sim$6\,kpc of the quasar centres (the angular resolution of the observations)

\subsection{Star formation rate limits}
\label{sec:sim}
In the sources without detected narrow H$\alpha$ emission, upper limits on the star-formation rates may be derived via a series of simulations.  We add three-dimensional Gaussian profiles to the data cubes (in RA, Dec and velocity space) with spatial widths that match the seeing of the observations (the broad H$\alpha$ PSF spatial profile) and a velocity width of $\sigma$=100\,kms$^{-1}$. The input profiles are placed at the quasar centres and the minimum intensity necessary to produce a 'detection', using our search procedure, is determined. In Table \ref{tab:sfr_in} we quote the resulting upper limits of the star formation rates, assuming the star forming region lies within the PSF region. In our test we assume velocity widths of $\sigma$=100\,kms$^{-1}$ for out input narrow H$\alpha$ emission. We note that to place limits on narrow emission with widths of $\sigma$=200\,kms$^{-1}$ the $\sigma$=100\,kms$^{-1}$ limits would need to be scaled up by a factor of $\sqrt 2$ to take into account the doubling in the velocity range of the noise calculation.   In two sources the limits on the star formation rates are $<$50\,$\rm M_{\odot}$yr$^{-1}$ indicating that these systems may contain very little star formation.

For ULASJ1234+0907 \citep{Banerji14a} we have also performed narrow-band H$\alpha$ imaging observations (detailed in Appendix \ref{sec:1234}). Above, we find that star formation within the PSF ($<$6.1\,kpc)  must be $<$4600\,$\rm M_{\odot}$yr$^{-1}$, consistent with the conclusion from the narrow band imaging, that the star formation must be within the central 8\,kpc.

\section{Discussion}
\label{sec:discussion}
We find evidence for significant star formation in 16 of 28 reddened quasars ($\simeq$54\,per cent) with an average  H$\alpha$ star formation rate of 320$\pm$70\,$\rm M_{\odot}$yr$^{-1}$ \footnote{ The star formation rates we derive assume that the extinctions towards the star formation regions are the same as towards the quasars, however, the extinctions may be smaller at locations offset from the quasar nucleus hence the extinction-corrected star formation rates could be over-estimated.}. In the remaining quasars, our observations provide limits on the H$\alpha$ star formation rates. In two of the sample star formation rates of 50\,$\rm M_{\odot}$yr$^{-1}$ or lower are ruled out within the central $\sim$6\,kpc. High-redshift starburst galaxies have typical star formation rates of $\sim$100\,$\rm M_{\odot}$yr$^{-1}$, therefore our limits could suggest that such levels of star formation have already been quenched in a fraction of the reddened quasar population. Alternatively, the star formation may be distributed over wider areas than our (seeing-determined) $\simeq<$6\,kpc resolution, thus eluding discovery in the procedure we have adopted. For the remaining eight quasars the limits are not as constraining ($<$110\,$\rm M_{\odot}$yr$^{-1}$ to $<$4600\,$\rm M_{\odot}$yr$^{-1}$) therefore it is possible that there is significant star formation within the central $\sim$6\,kpc of these quasars. Our findings are consistent with the $\sim$71\,per cent of quasar host galaxies found to be actively forming stars in a sample of X-ray selected obscured quasars at z$\sim$2 \citep{Mainieri11}.

Observations of high-redshift unobscured luminous quasars, made at far infrared and submillimetre wavelengths, have shown that approximately a quarter of the unobscured sample are bright in the far infrared and therefore are likely associated with recent star formation \citep[e.g.][]{Omont03,Priddey03}. A higher fraction of the reddened quasars contain star formation compared to similarly luminous blue quasars, although a direct comparison is not possible because different star formation indicators are used for the two populations. {\it Hubble Space Telescope} observations of z$\sim$1-2 unobscured luminous quasars have been used to estimate mean star formation rates of $\sim$50-350\,$\rm M_{\odot}$yr$^{-1}$ \citep{Floyd13}, indicating that the reddened quasars exhibit higher levels of star formation than the unobscured quasars.

The presence of high star formation rates, associated with larger dust extinctions, provides a natural explanation for the observed correlation between luminous reddened quasars and the presence of significant star formation. Far-infrared to millimetre dust continuum observations have already constrained star formation rates in three of our reddened quasars: $\sim$2000\,$\rm M_{\odot}$yr$^{-1}$ for ULASJ1234+0907 \citep{Banerji14a}, $\sim$275\,$\rm M_{\odot}$yr$^{-1}$ for ULASJ1002+0137 \citep{Brusa14} and $>$250-1600\,$\rm M_{\odot}$yr$^{-1}$ for ULASJ1539+0557 \citep{Feruglio14}.  The average star formation rate of 320$\pm$70\,$\rm M_{\odot}$yr$^{-1}$ we measure is consistent with the rates deduced from far-infrared and millimetre dust continuum observations. The high level of star formation observed in our sample is consistent with the hypothesis that the reddened quasars represent a post-merger stage in the evolution of massive galaxies in which the merger has induced both a starburst and increased fuelling of the central supermassive black hole \citep[e.g][]{Hopkins08}.

We now draw comparisons between the typical scales of the star-forming regions seen in our reddened quasars and those measured in other populations of star-forming galaxies and AGN.

\subsection{Comparison to SMGs at z$\sim$2}
In the evolutionary sequence proposed by \cite{Hopkins08}, merger induced star formation triggers accretion onto the central black-holes of the galaxies such that the systems appear as dusty (reddened) quasars. The dusty star forming galaxies of the SMG population may then represent a phase that is followed by a reddened quasar phase. The average star formation rate we measure in the reddened quasar sample of 320$\pm$70\,$\rm M_{\odot}$yr$^{-1}$ matches the average H$\alpha$-derived star formation rate of the z$\sim$2 SMGs studied in \cite{Alaghband-Zadeh12} of 370$\pm$90\,$\rm M_{\odot}$yr$^{-1}$ (Fig. \ref{fig:sfr_r}) indicating that powerful starbursts are occurring in the reddened quasars, similar to the SMGs. 

In the z$\sim$2 SMGs, star formation observed through H$\alpha$ studies \citep{Swinbank06, Alaghband-Zadeh12, Menendez-Delmestre13} is seen to be extended on scales of $\sim$2-17\,kpc, leading to the proposal that the SMGs consist of multiple merging components. Furthermore, in the AGN-SMG system (SMMJ0217-0503; \citealt{Alaghband-Zadeh12}) there is an offset of $\sim$18\,kpc between the AGN and the star-forming SMG component.   We find evidence for extended star formation on scales of $\sim$8\,kpc in five of our sample, consistent with the extended star formation observed in the SMGs. In the majority of our sample the star forming regions are constrained to lie within $\sim$6\,kpc of the quasar centres and therefore are at the lower end of the range of offsets in SMGs (Fig. \ref{fig:sfr_r}). It is possible that the reddened quasar phase represents a more advanced stage in a merger sequence, when the star forming gas is more compact than observed in the SMGs. However, a comparison between the star-forming properties of the reddened quasar host galaxies and SMGs would ideally be undertaken for sub-populations of comparable total mass.

In the hydrodynamic simulations of SMG mergers detailed in \cite{Narayanan09}, the star formation rate during the first passage and inspiral phases of the merger is $\sim$200\,$\rm M_{\odot}$yr$^{-1}$ (rising up to $\sim$1000\,$\rm M_{\odot}$yr$^{-1}$ at final coalescence) matching the average star formation rates of our reddened quasar sample (320$\pm$70\,$\rm M_{\odot}$yr$^{-1}$). In the simulations, the sizes of the $^{12}$CO(3-2) emission regions, tracing the molecular gas and therefore the fuel for the star formation, are predicted to be $\sim$1.5\,kpc during the star burst phase and only extend up to $\sim$2.5\,kpc for cases where the intense star burst occurs at the early stages of the merger (Fig. \ref{fig:sfr_r}). The origin of the compact star formation in the simulations is the gas being pushed into the centre during the merger. While the sizes are consistent with the upper limits constrained from our observations, the simulations are tracing the molecular gas which fuels star formation, whereas our H$\alpha$ emission observations arise from ionized gas following the formation of massive stars.

\subsection{Comparison to low redshift reddened quasars}
\cite{Urrutia08} present {\it Hubble Space Telescope} observations of low-redshift reddened quasars from the FIRST/2MASS Red Quasar Survey \citep{Glikman07} finding that they exhibit multiple, interacting components and tidal tails, consistent with an origin in recent merger activity. Their sample is composed of systems with a wide range of separations, suggesting that the reddened quasar phase in the low-redshift sample may occur at different stages of the merger process; from early on, where the merging galaxies are still identifiable, through more advanced stages, where multiple nuclei are observed, as well as in the final coalescence stage when tidal tails are visible.

Five of the low-redshift reddened quasar sample exhibit more than one component, with separations of 1.2$-$9.2\,kpc, and starburst-like star formation rates of $\sim$100\,$\rm M_{\odot}$yr$^{-1}$ (calculated using the far-infrared luminosities from \citealt{Urrutia12} and \citealt{Kennicutt12}) as shown in Fig. \ref{fig:sfr_r}. Our finding that star formation is located within $\sim$6\,kpc of the quasar nuclei in the majority of the high-redshift reddened quasar sample is consistent with the component separations of these clumps (associated with the quasar hosts) in the low-redshift sample. The similarity in the spatial scales of the high rates of star-formation provides some support for associating the high-redshift reddened quasars with the late stage (final coalescence) of a galaxy merging event, as also found by \cite{Glikman15}.

\subsection{Comparison to high redshift quasars}
Observations of the star formation rates and locations in unobscured quasars of comparable luminosity to our reddened quasars are few in number. In a study of five $z\sim 2$ quasars (\citealt{Vayner14}; Fig. \ref{fig:sfr_r}), two objects are found to possess star formation rates of $\sim$40-110\,$\rm M_{\odot}$yr$^{-1}$ at distances of 2-4\,kpc from the quasars, with three non-detections. We therefore find that a significant fraction of the reddened quasars are associated with more extreme star formation rates than is common for unobscured quasars, although the spatial locations of the events may be consistent. \cite{Walter09} show that at $z\sim$6, an extreme star formation rate (1700\,$\rm M_{\odot}$yr$^{-1}$) in a quasar host galaxy is confined to the central $\sim$1\,kpc (Fig. \ref{fig:sfr_r}) which they suggest may be the result of a merger, although they do not rule out the possibility of infall of cold gas from streams triggering the star formation \citep{Dekel09a}. The limits on the sizes of the majority of our sample are not as constraining as the compact star formation observed in this source, however we cannot rule out that a similar origin may apply for our sample.

\subsection{Feedback}
IFU observations enable both the spatial locations of star forming regions to be determined and the effects of the AGN on the host galaxy to be studied. Recent work supports a mixture of both positive and negative AGN-feedback affecting the host galaxy star formation. More specifically, AGN outflows have been suggested both to cause gas to condense to form stars and also remove (or heat) gas and thus inhibit star formation \citep[e.g.][]{Silk98,Silk05,Zubovas13,Farrah12,Zinn13}.

\cite{Ishibashi13} suggest that star formation can occur in AGN-feedback driven outflows and that the AGN-induced star formation can explain a number of other galaxy properties such as the evolution of their size and structure \citep{Ishibashi14}. In general, models predict significant influence of an AGN-driven flow close to the AGN, whereas there is much less clarity regarding the significance of any AGN-induced activity on large scales.  The models of \cite{Nayakshin14} predict positive-feedback due to the effect of the AGN on star formation in the early gas-rich phases of galaxy evolution with negative feedback important on much larger scales at later times. Our observations of star formation found to lie in regions of $\sim$6\,kpc around the quasar centre indicate that it is possible that the AGN has induced star formation in these systems on these scales. However, higher resolution (likely AO assisted) IFU observations are required to resolve the morphology and dynamics of the star forming regions much closer to the quasar nuclei ($<\sim$1-3\,kpc) such that AGN-induced star formation could be identified or discounted. Positive AGN-feedback occurring at large offsets from the nucleus has been suggested from observations of extended Ly$\alpha$ around a quasar at $z\sim$3 \citep{Rauch13}. We find extended star formation in five of our sample, on $\sim$8\,kpc scales, suggesting that AGN-driven outflows could trigger star formation on these scales. 

The detection of kiloparsec scale molecular outflows in both high redshift and local quasars \citep{Feruglio10, Cicone12,Maiolino12,Cicone14} indicates that outflows could affect star formation in host galaxies on the $\sim$5\,kpc scales of our star-formation detections. Both the resolved observations and the interpretation are however not straightforward \citep[e.g.][]{Rupke11}, particularly if both positive and negative feedback occur with different spatial and temporal dependencies. The large average star formation rate we measure for 16 quasars within $\sim$6\,kpc does not support strong negative feedback on such scales in approximately half our sample. However negative feedback may be responsible for the two sources in which the star formation rates are constrained to be $<$50\,$\rm M_{\odot}$yr$^{-1}$ suggesting very little star formation is occurring in these host galaxies. The limits on the star formation rates in the remaining eight quasars are not constraining enough to rule out significant star formation therefore feedback effects cannot be assessed.

\cite{Cresci14} combine their detections of narrow H$\alpha$ emission in ULASJ1002+0137 (see Appendix \ref{sec:1002}) with [OIII] and HST $U$-band observations and propose that ULASJ1002+0137 contains both positive and negative feedback. They detect an outflow in [OIII] extending up to 13\,kpc from the AGN; the AGN appears to reside in a region with little star formation, surrounded by areas experiencing elevated star formation. Their interpretation invokes an outflow, which removes gas within the extent of the outflow, therefore inhibiting star formation, while also inducing star formation at the interface between the outflow and gas in the host galaxy. A similar scenario is seen around a z$\sim$2 quasar \citep{Cano-Diaz12} in which extended star formation (on scales of several kpc; Fig. \ref{fig:sfr_r}) is observed, however, there is no star formation observed in the region where a strong outflow is detected in [OIII]. To explore the various feedback effects further, and determine the fraction of reddened quasars which exhibit such a geometry, a comparison to the resolved [OIII] emission in each source is required.  Complimentary [OIII] observations can be used to compare the outflows and star formation morphologies. Higher-resolution observations of multiple diagnostic emission lines will enable the gas properties and kinematics to be studied, allowing the importance of AGN-feedback in the evolution of the host galaxies to be constrained.

\section{Conclusions}
We have utilised IFU observations to study the star formation properties of a sample of 28 reddened quasars at $z\sim2$, finding evidence for star formation in 16 reddened quasars (i.e. 57\%), with an average star formation rate of 320$\pm$70\,$\rm M_{\odot}$yr$^{-1}$. The star forming regions detected are constrained to lie within $\sim$6\,kpc of the central AGN in the majority of the sample, however in five of the sample there is evidence for resolved, extended emission around the quasar centre on scales of $\sim$8\,kpc. In two of the sources in which we do not detect star formation the star formation rates are constrained to be $<$50\,$\rm M_{\odot}$yr$^{-1}$ suggesting very little star formation is occurring in these host galaxies. However, the limits on the star formation rates for the remaining eight quasars are not as constraining, implying that significant star formation may be present in the majority of our sample. Higher resolution observations of the star formation in the reddened quasar sample are required to disentangle the potential positive and negative feedback effects of the AGN on the star formation properties of its host galaxy.

\section{Acknowledgements}
\label{sec:ackn}
The authors thank the anonymous referee for their helpful and detailed comments on the paper. The authors also thank Joe Hennawi for useful discussions. The authors thank the Science and Technology Facilities Council (STFC) via the Consolidated Grant awarded to the Institute of Astronomy, Cambridge. SA-Z acknowledges the support from Peterhouse, Cambridge, and the Institute of Astronomy, Cambridge. The narrow-band imaging study is based on observations made with ESO Telescopes at the La Silla Paranal Observatory under programme ID: 290.A-5062. The IFU study is based on observations made with ESO Telescopes at the La Silla Paranal Observatory under programme IDs:383.A-0573 (PI:McMahon) and 091.A-0341 (PI:Banerji).

\bibliographystyle{mn2e}

\bibliography{refs_short}

\appendix

\section{ISAAC narrow-band H$\alpha$ imaging of ULASJ1234+0907}
\label{sec:1234}
For ULASJ1234+0907 we have also performed narrow-band imaging observations using the ISAAC (Infrared Spectrometer And Array Camera) camera on the VLT \citep{Moorwood98}. We used a narrow-band filter at 2.29\,$\mu$m corresponding to the observed wavelength of H$\alpha$ in this quasar. Observations were carried out in 2013 March. The ISAAC field of view is 2.5$\times$2.5\,arcmin and the pixel scale is 0.148\,arcsec per pixel. ULASJ1234+0907 is the most extreme object in our sample. With ($J-K$)=5.3 (Vega), it is by far the reddest quasar currently known and the reddest near infra-red luminous source in the extragalactic sky.  The host galaxy of the rapidly accreting black-hole is a hyperluminous starburst with an enormous star formation rate conservatively estimated to be $\sim$2000$\rm M_{\odot}$yr$^{-1}$ \citep{Banerji14a}.  The objective of the narrow-band observations was therefore to detect extended H$\alpha$ emission associated with star formation in the quasar host galaxy. A broad $K$-band continuum image was also taken using ISAAC to allow calibration of the narrow-band image and subtraction of continuum emission from any narrow-band detected sources. The narrow-band observations were split into two Observing Blocks (OBs) for ease of scheduling. Each OB consisted of seven dithered exposures of 4$\times$75s each resulting in a total integration time of 70 minutes across the two OBs. The $K_S$-band observations were taken using 15 dithered exposures of 6$\times$10s each resulting in a total integration time of 15 minutes. All observations were carried out under photometric conditions and in very good seeing of $<$0.5\,arcsec in order to be able to spatially resolve kiloparsec scale structures associated with extended star formation in the quasar host.    

The data was reduced using standard ESO tools provided as part of the \textit{gasgano} package. Data reduction steps included dark subtraction, flat-fielding and co-addition of the individual exposures to produce the combined images. Images were registered onto the World Coordinate System (WCS) using the positions of known 2MASS stars in the field. Flux calibration was performed using standard stars observed in the $K_S$ filter on the same night. The median $K_S$ band zero-point derived from these observations is 24.16. We applied an offset of 1.9 mags \citep{Retzlaff10} in $K_S$ to convert the magnitudes to the AB system. The relative throughput between the broad and narrow-band filters was determined using a single 2MASS star in our field. Due to the lack of other bright objects in the field, this flux calibration is expected to be accurate at the level of 0.1 mags. As the two narrow-band OBs were executed on different nights, we reduced each of these separately to produce two co-added narrow-band images. After ensuring that the photometry and astrometry were consistent between the two images, the two reduced narrow-band images were then combined to produce our final narrow-band image. 

The measured FWHM of the quasar in the $K_S$-band is 0.41\,arcsec which is consistent with the size of the seeing disk of the ISAAC observations. The FWHM of the quasar in the two narrow-band images is 0.43\,arcsec and 0.51\,arcsec once again consistent with the seeing of the ISAAC observations.

The quasar appears to be unresolved in both the broad and narrow-band observations. In order to search for faint extended H$\alpha$ emission, we attempted to perform PSF subtraction on the quasar image using the single bright star in the field. The star was subtracted from the quasar images after scaling the flux of the star to match that of the quasar. Although some residuals remained due to imperfect subtraction, the PSF subtracted images are consistent with noise and we do not find any evidence for obvious extended H$\alpha$ emission around the quasar. 

In order to derive the noise on the H$\alpha$ images, we extracted a region of 256$\times$256 pixels around the quasar in which there are no other bright sources. The quasar broad-line emission was then masked from the images by removing all pixels above 5$\sigma$ which are all located within the seeing disk of the image. The remaining pixels were then used to calculate the RMS noise properties of the narrow-band images. The RMS noise is 0.10\,$\mu$Jy per pixel in the combined narrow-band image which translates to an H$\alpha$ flux of  1.7$\times$10$^{-18}$\,erg/s/cm$^2$ per pixel.

\begin{figure}
\centering
\includegraphics[scale=0.6,angle=0]{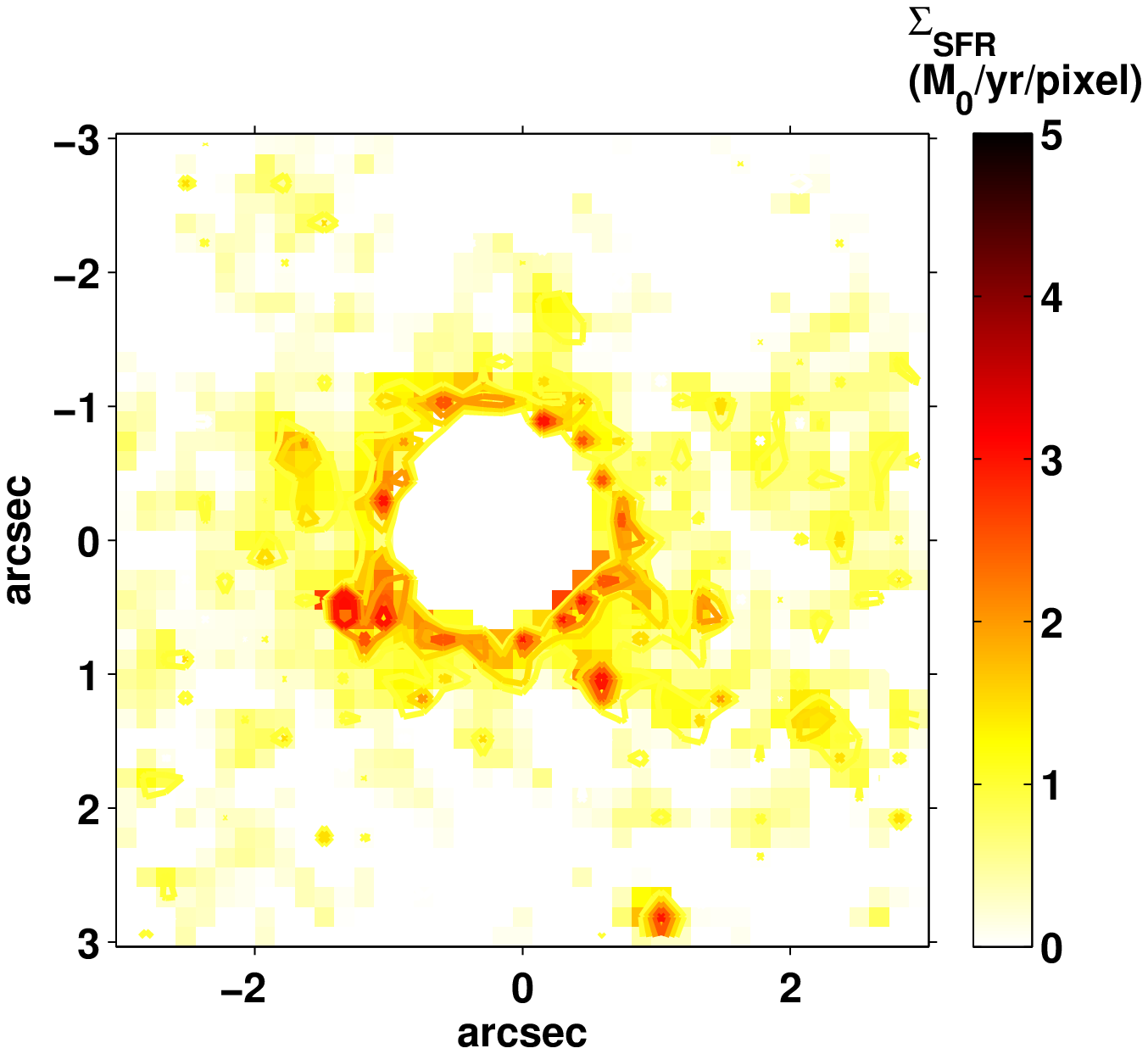}
\caption{Narrow-band H$\alpha$ emission map in a region of 6$\times$6\,arcsec around ULASJ1234+0907 where the H$\alpha$ emission per pixel has been converted to a star formation rate density per pixel. The pixel size is 0.148\,arcsec. The image has been smoothed using median filtering over a 2$\times$2 pixel box. The 1$\sigma$ contours can also be seen. From this map, we obtain a 1$\sigma$ limit on the star formation rate density of $\Sigma_{\rm{SFR}}$ $<$1\,M$_\odot$yr$^{-1}$ per pixel}
\label{fig:sfrdens}
\centering
\end{figure}

We subtract the broad-line H$\alpha$ emission from our narrow-band image such that we are only sensitive to extended star formation at radii outside the seeing FWHM ($\sim$1\,arcsec from the quasar nucleus). This corresponds to physical scales of $\gtrsim$8\,kpc at $z=2.503$.  At scales of $>$8\,kpc from the quasar, our narrow-band imaging allows us to place limits on the star formation rate density (converting H$\alpha$ flux to a star formation rate using \cite{Kennicutt12}) of $\Sigma_{\rm{SFR}}$ $<$3\,$\rm M_{\odot}$yr$^{-1}$ per pixel or $<2$\,$\rm M_{\odot}$yr$^{-1}$kpc$^{-2}$ (3$\sigma$). In Fig. \ref{fig:sfrdens} we show the star formation rate density map in a box of 6$\times$6\,arcsec around ULASJ1234+0907. The nuclear source has been masked out from the image. 

The total star formation taking place in ULASJ1234+0907 has been constrained as $\sim$2000\,$\rm M_{\odot}$yr$^{-1}$ from far infra-red observations \citep{Banerji14a}, enabling a comparison to the limits on the level of unobscured star formation derived from our H$\alpha$ narrow-band imaging. Given the extreme starburst luminosity of ULASJ1234+0907 and even accounting for a dust-extinction of $A_V=6$ mags as measured towards the central black-hole \citep{Banerji12}, the H$\alpha$ flux due to star formation is expected to be 3.4$\times$10$^{-17}$\,erg/s/cm$^2$. This is a lower limit given that the extinction may be less severe towards the star-forming regions. Assuming this star formation is occurring over a region of $\sim$6 kpc or 0.75\,arcsec consistent with observations of SMGs \citep{Swinbank06, Alaghband-Zadeh12, Menendez-Delmestre13}, this means that the H$\alpha$ flux per pixel due to star formation is expected to be $\sim$7$\times$10$^{-18}$\,erg/s/cm$^2$. The noise in our co-added narrow-band image is 1.7$\times$10$^{-18}$\,erg/s/cm$^2$ per pixel so we should have detected this star formation at the $\sim$4$\sigma$ level if it was occurring at radii $>$8\,kpc from the central black-hole. We conclude therefore that the starburst in ULASJ1234+0907 is located within $\sim$8\,kpc of the central black-hole.  In Section \ref{sec:results} we find that any star formation within the PSF ($<$6.1\,kpc) must have a star formation rate $<$4600\,$\rm M_{\odot}$yr$^{-1}$. These results are therefore consistent with the findings of the narrow band imaging that the star formation  must be contained within the central 8\,kpc.

\section{ULASJ1002+0137}
\label{sec:1002}
Using the same H$\alpha$ data, \cite{Cresci14} highlight two regions (at 9$\sigma$ and 4$\sigma$ significance levels) of narrow H$\alpha$ emission in ULASJ1002+0137 from which they measure a combined  H$\alpha$ star formation rate of $\sim$230\,$M_{\odot}$/yr,  consistent with the  star formation rate of $\sim$275\,$M_{\odot}$/yr constrained from far-infrared observations \citep{Brusa14}. The narrow H$\alpha$ profile of ULASJ1002+0137 presented in both \cite{Bongiorno14} and \cite{Cresci14} is present in our integrated spectrum (see Paper I).  A strong sky-emission line is coincident with the narrow H$\alpha$ emission.  Using our noise-weighted, broad profile-subtracting, detection scheme we detect a smaller narrow H$\alpha$ signal with a star formation rate of 47$\pm$9\,$M_{\odot}$/yr found to lie in a region of  $\sim$11\,kpc around the quasar centre. Owing to the 0.8$''$ seeing of the observations we do not attempt to undertake a spatially resolved analysis of this object. We note that our procedure is tailored for use on multiple objects with no prior information available about the shape of both the broad and narrow H$\alpha$ lines.

\onecolumn
\section{Figures}
\label{app:figs}

\begin{figure}
\includegraphics[width=0.95\textwidth]{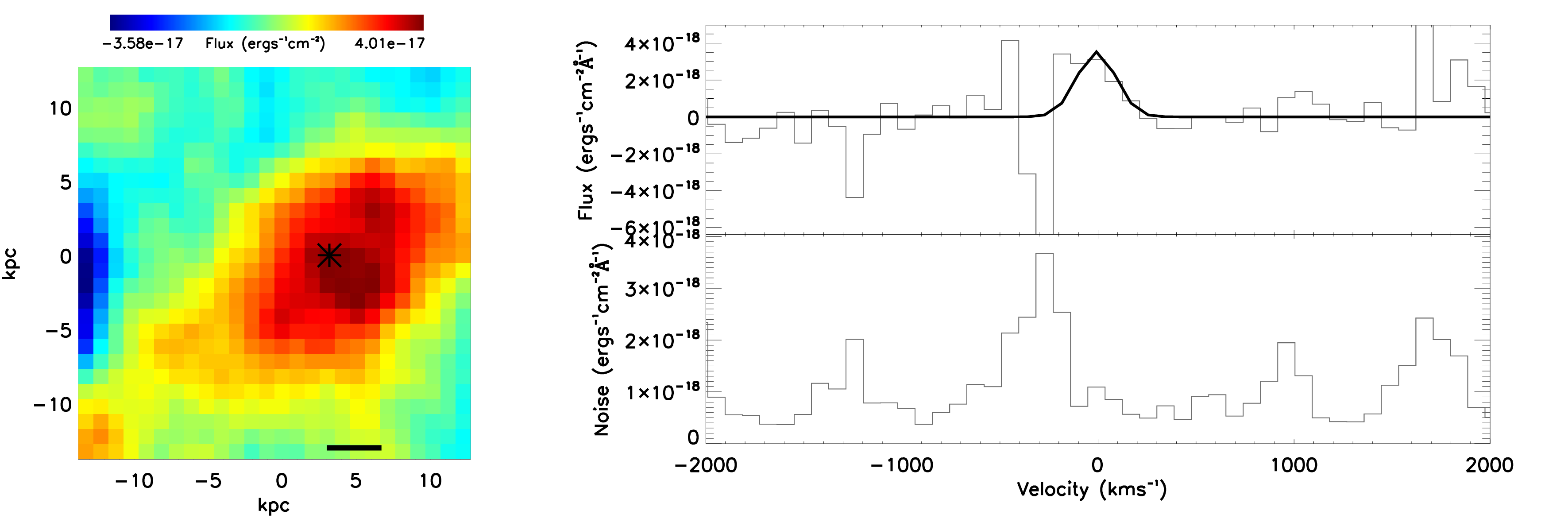}
\caption{Left: Map of the narrow H$\alpha$ emission in ULASJ1002+0137, created by integrating the profile-subtracted cube from the centre of the narrow line $\pm$150\,kms$^{-1}$ (1.5$\sigma$). The profile-subtracted cube is made by subtracting the broad line profile (scaled using the maximum likelihood technique) from the spectrum in each binned spaxel. Since we find evidence for extended (spatially resolved) narrow H$\alpha$ emission in this case the spaxels are binned such that each spaxel represents the sum of the surrounding spaxels in a region the size of which maximises the S/N of the detection. The black cross marks the quasar centre and corresponds to the aperture in which the total spectrum has evidence for a narrow Gaussian of width $\sigma$=100\,kms$^{-1}$.  We show the FWHM of the PSF-aperture by the black line in the bottom, right-hand corner. Top Right: The total spectrum from the aperture at the quasar centre with the scaled Gaussian overlayed. Bottom Right: The noise spectrum (as detailed in Section \protect\ref{sec:noise}).}
\label{fig:narrow_ULASJ1002+0137}
\end{figure}

\begin{figure}
\includegraphics[width=0.95\textwidth]{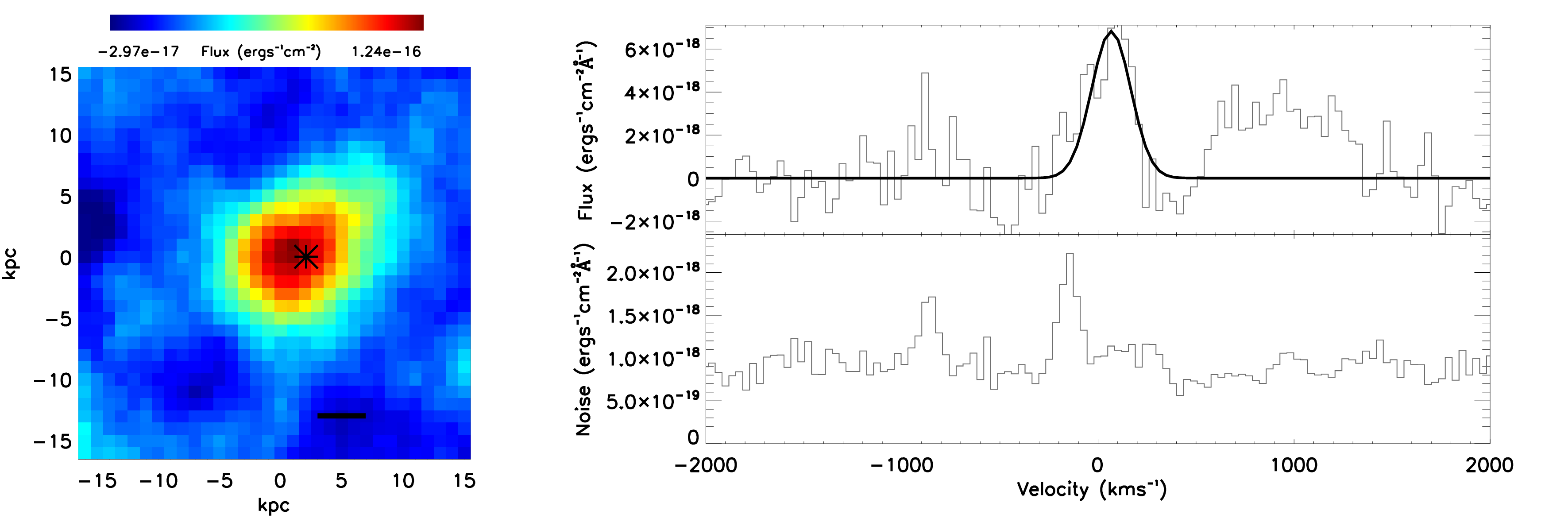}
\caption{As for \protect\ref{fig:narrow_ULASJ1002+0137} for the narrow H$\alpha$ emission in VHSJ1556-0835. Since the narrow H$\alpha$ emission is spatially unresolved the spaxels are binned such that each spaxel represents the sum of the surrounding spaxels in a region the size of PSF-aperture. The black cross marks the quasar centre and corresponds to the PSF-aperture in which the total spectrum has evidence for a narrow Gaussian of width $\sigma$=100\,kms$^{-1}$. We show the FWHM of the PSF-aperture by the black line in the bottom, right-hand corner.}
\label{fig:narrow_VHSJ1556-0835}
\end{figure}

\begin{figure}
\includegraphics[width=0.95\textwidth]{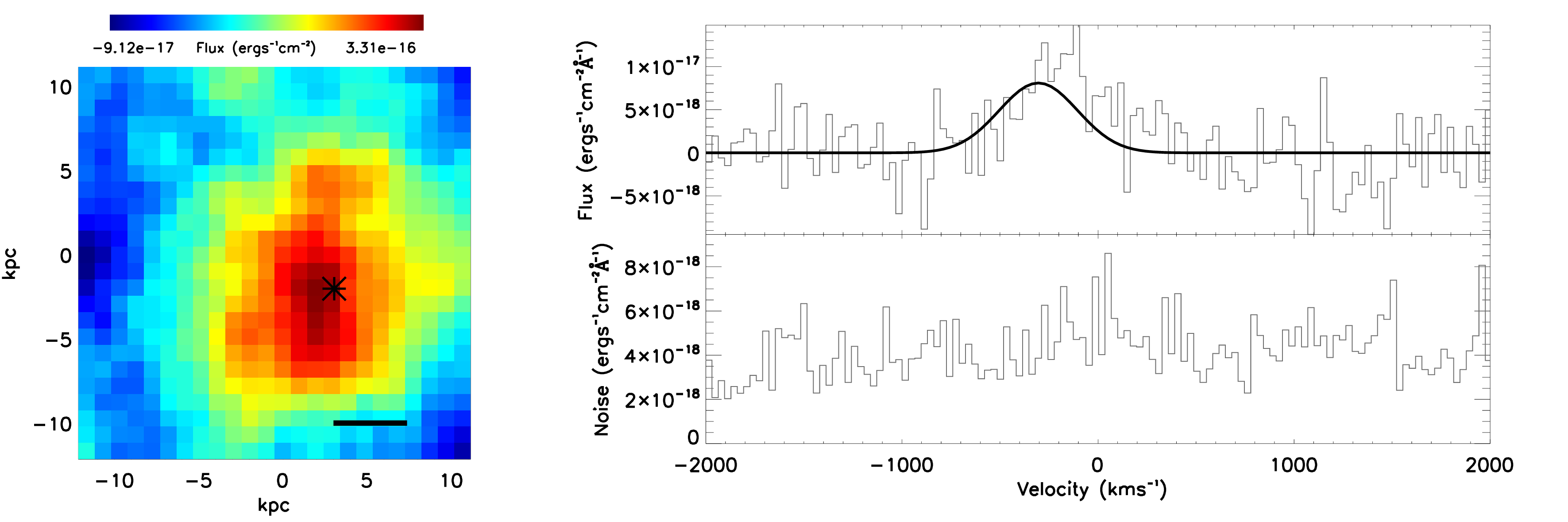}
\caption{As for \protect\ref{fig:narrow_ULASJ1002+0137} for the resolved narrow H$\alpha$ emission in VHSJ2028-4631. A narrow Gaussian profile with a width of $\sigma$=200\,kms$^{-1}$ was found to give a better fit therefore the map is created by integrating over $\pm$300\,kms$^{-1}$ (1.5$\sigma$) and the Gaussian profile overlayed has a width of $\sigma$=200\,kms$^{-1}$.}
\label{fig:narrow_VHSJ2028-4631}
\end{figure}

\begin{figure}
\includegraphics[width=0.95\textwidth]{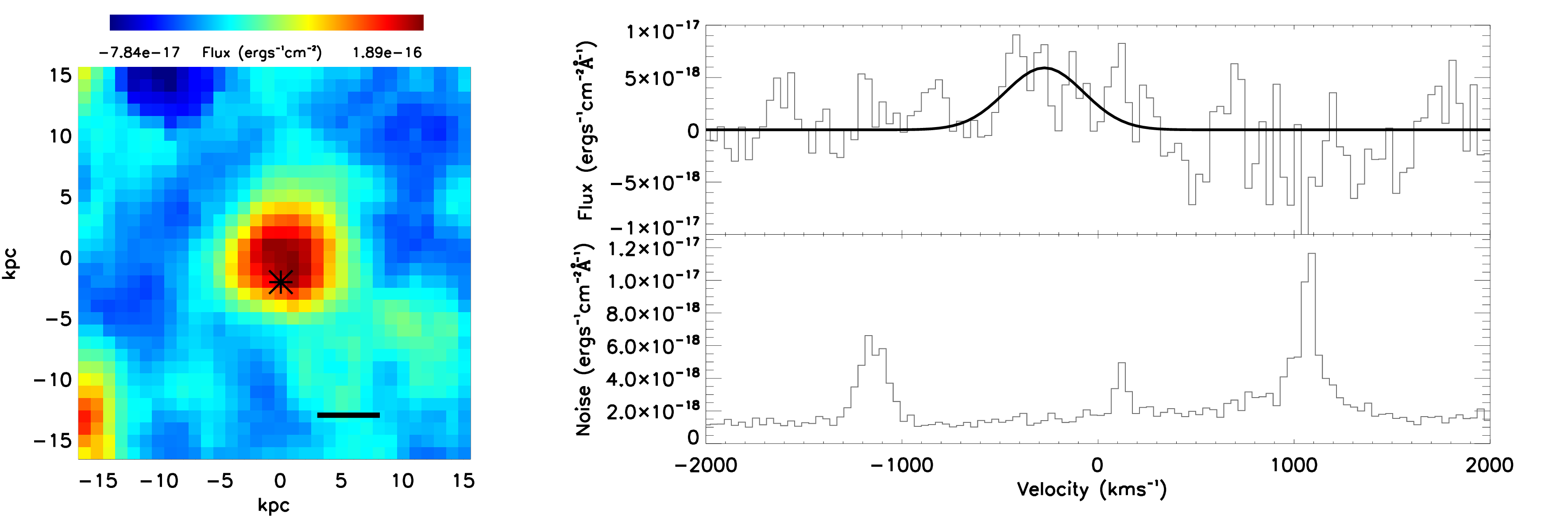}
\caption{As for \protect\ref{fig:narrow_VHSJ1556-0835} for the unresolved narrow H$\alpha$ emission in VHSJ2028-5740. A narrow Gaussian profile with a width of $\sigma$=200\,kms$^{-1}$ was found to give a better fit therefore the map is created by integrating over $\pm$300\,kms$^{-1}$ (1.5$\sigma$) and the Gaussian profile overlayed has a width of $\sigma$=200\,kms$^{-1}$.}
\label{fig:narrow_VHSJ2028-5740}
\end{figure}

\begin{figure}
\includegraphics[width=0.95\textwidth]{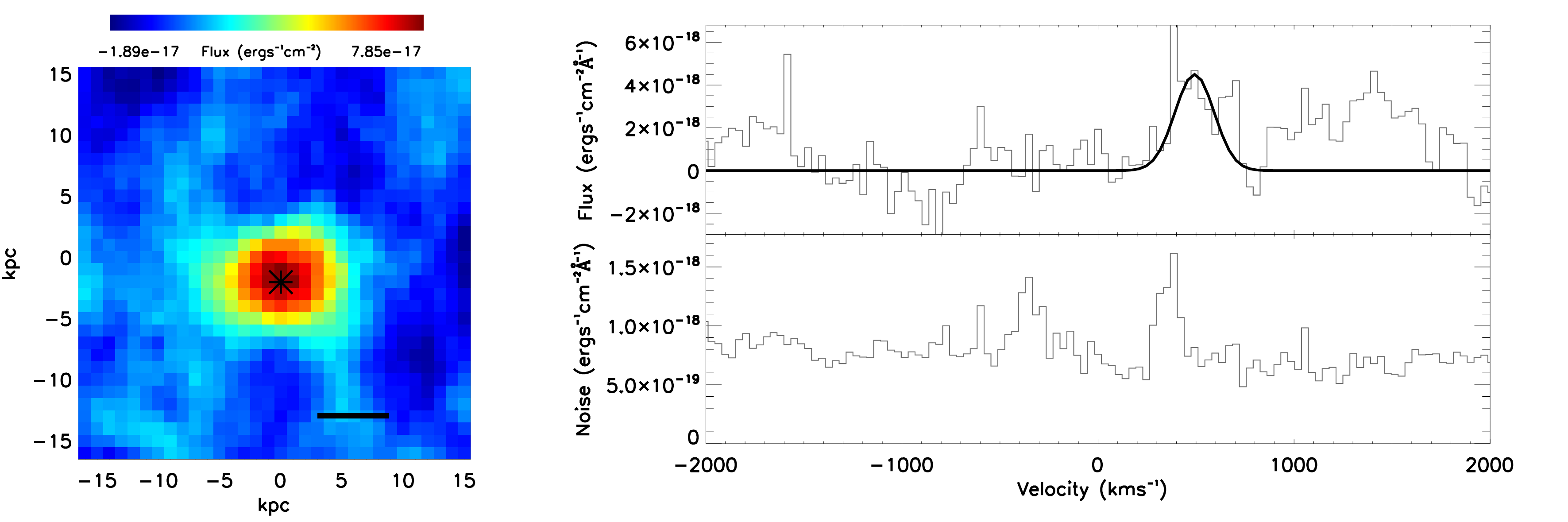}
\caption{As for \protect\ref{fig:narrow_VHSJ1556-0835} for the unresolved narrow H$\alpha$ emission in VHSJ2048-4644. }
\label{fig:narrow_VHSJ2048-4644}
\end{figure}

\begin{figure}
\includegraphics[width=0.95\textwidth]{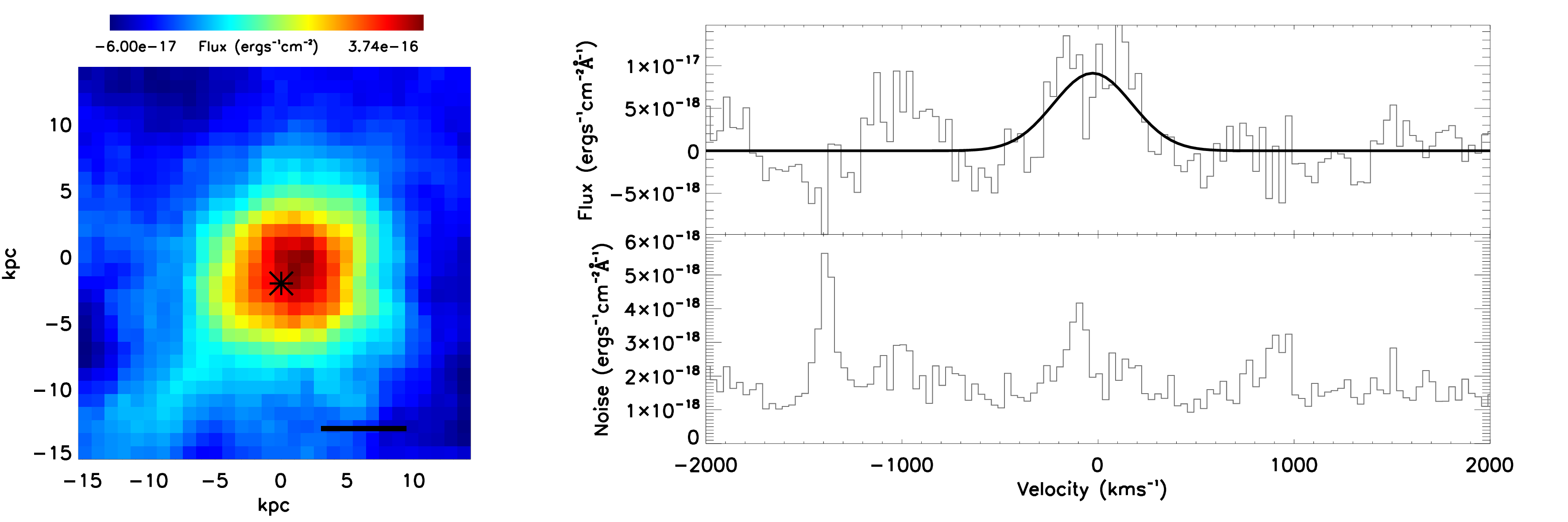}
\caption{As for \protect\ref{fig:narrow_VHSJ1556-0835} for the unresolved narrow H$\alpha$ emission in VHSJ2100-5820. A narrow Gaussian profile with a width of $\sigma$=200\,kms$^{-1}$ was found to give a better fit therefore the map is created by integrating over $\pm$300\,kms$^{-1}$ (1.5$\sigma$) and the Gaussian profile overlayed has a width of $\sigma$=200\,kms$^{-1}$.}
\label{fig:narrow_VHSJ2100-5820}
\end{figure}

\begin{figure}
\includegraphics[width=0.95\textwidth]{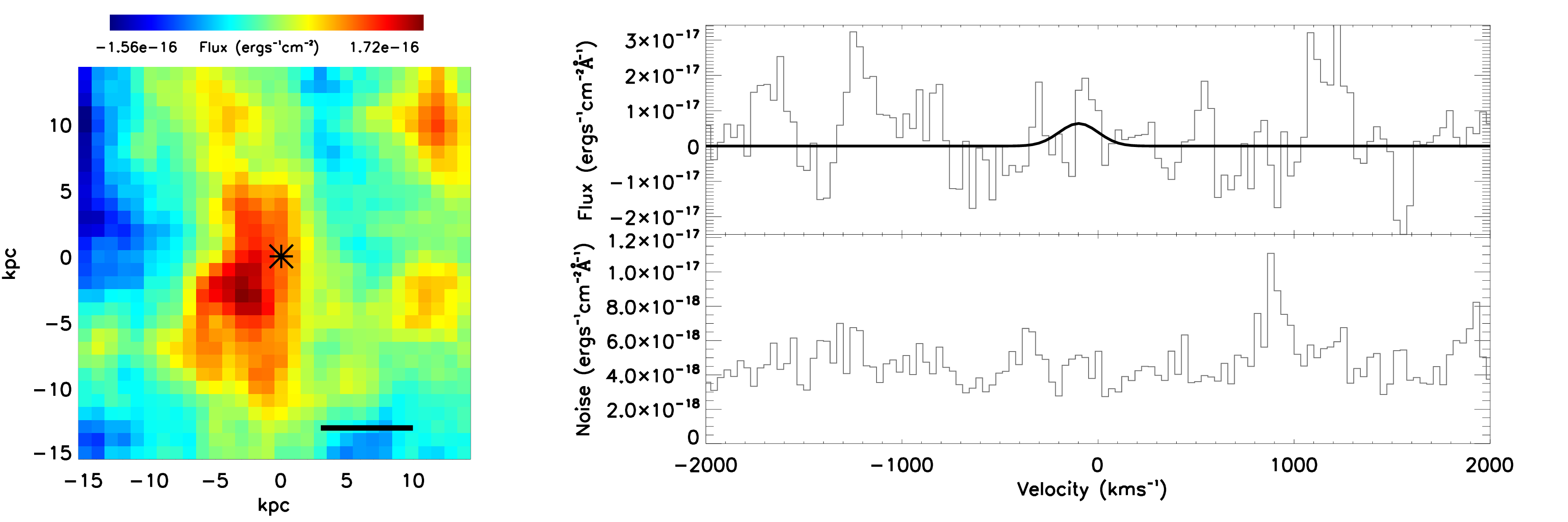}
\caption{As for \protect\ref{fig:narrow_VHSJ1556-0835} for the unresolved narrow H$\alpha$ emission in VHSJ2101-5943. This source is removed from the star formation sample as detailed in Section \protect\ref{sec:crit}.}
\label{fig:narrow_VHSJ2101-5943}
\end{figure}

\begin{figure}
\includegraphics[width=0.95\textwidth]{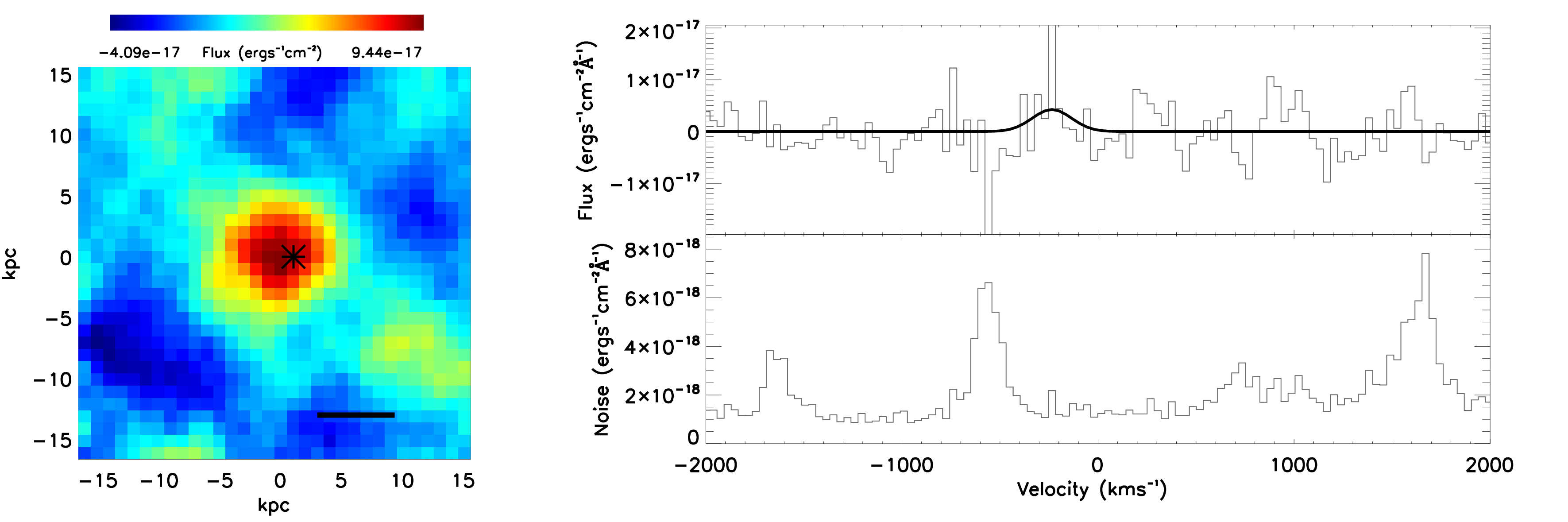}
\caption{As for \protect\ref{fig:narrow_VHSJ1556-0835} for the unresolved narrow H$\alpha$ emission in VHSJ2115-5913. }
\label{fig:narrow_VHSJ2115-5913}
\end{figure}

\begin{figure}
\includegraphics[width=0.95\textwidth]{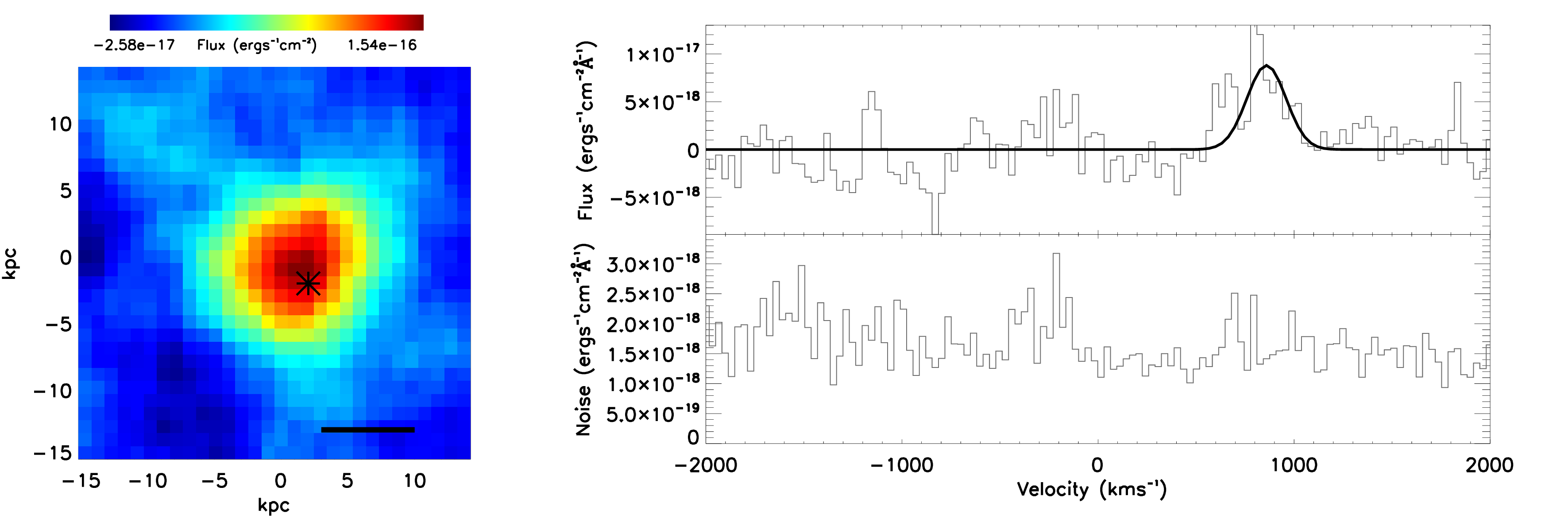}
\caption{As for \protect\ref{fig:narrow_VHSJ1556-0835} for the unresolved narrow H$\alpha$ emission in VHSJ2130-4930. }
\label{fig:narrow_VHSJ2115-5913}
\end{figure}

\begin{figure}
\includegraphics[width=0.95\textwidth]{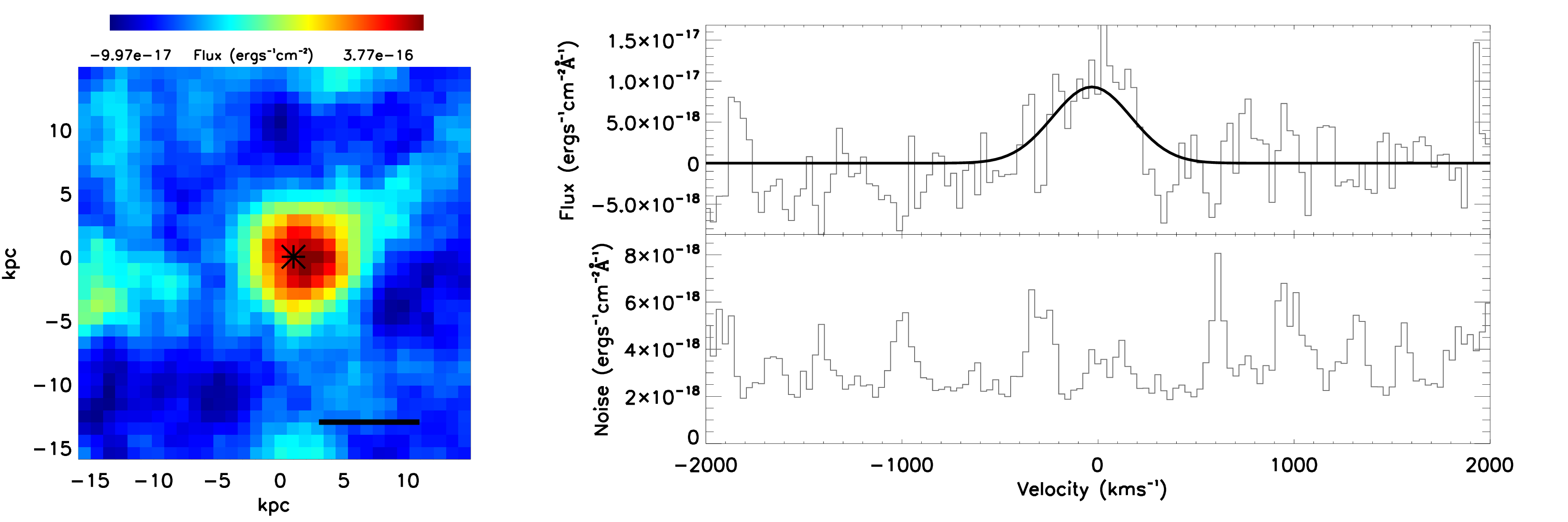}
\caption{As for \protect\ref{fig:narrow_VHSJ1556-0835} for the unresolved narrow H$\alpha$ emission in VHSJ2141-4816. A narrow Gaussian profile with a width of $\sigma$=200\,kms$^{-1}$ was found to give a better fit therefore the map is created by integrating over $\pm$300\,kms$^{-1}$ (1.5$\sigma$) and the Gaussian profile overlayed has a width of $\sigma$=200\,kms$^{-1}$. }
\label{fig:narrow_VHSJ2115-5913}
\end{figure}

\begin{figure}
\includegraphics[width=0.95\textwidth]{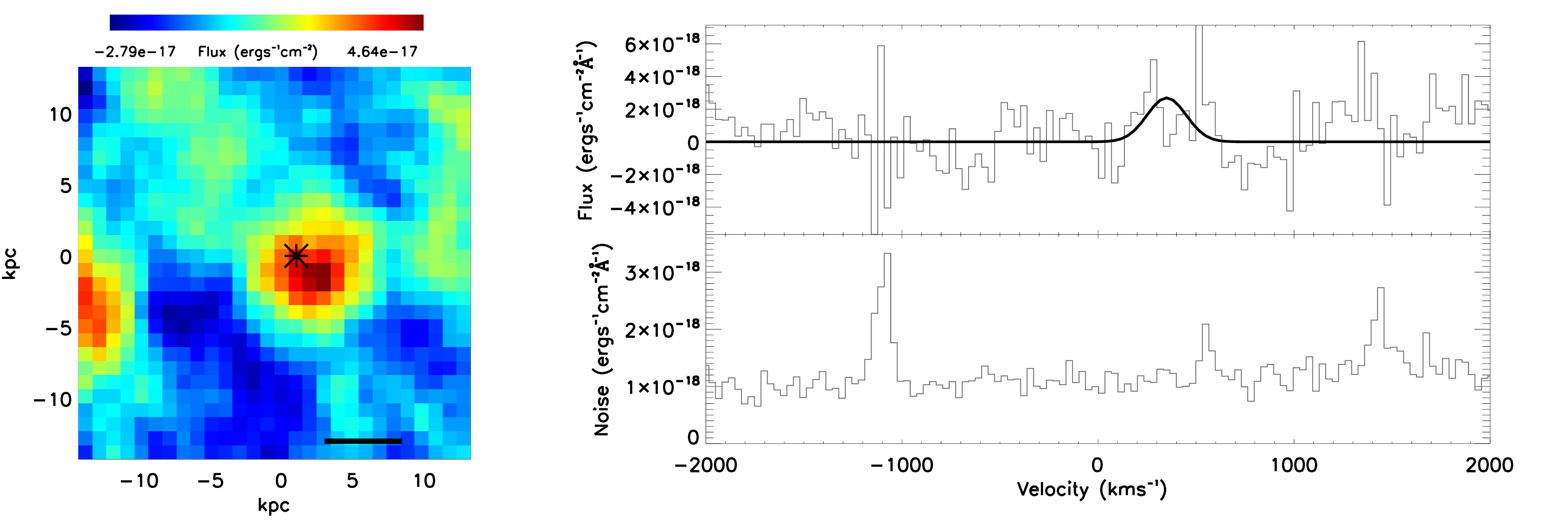}
\caption{As for \protect\ref{fig:narrow_VHSJ1556-0835} for the unresolved narrow H$\alpha$ emission in VHSJ2143-0643. }
\label{fig:narrow_VHSJ2143-0643}
\end{figure}

\begin{figure}
\includegraphics[width=0.95\textwidth]{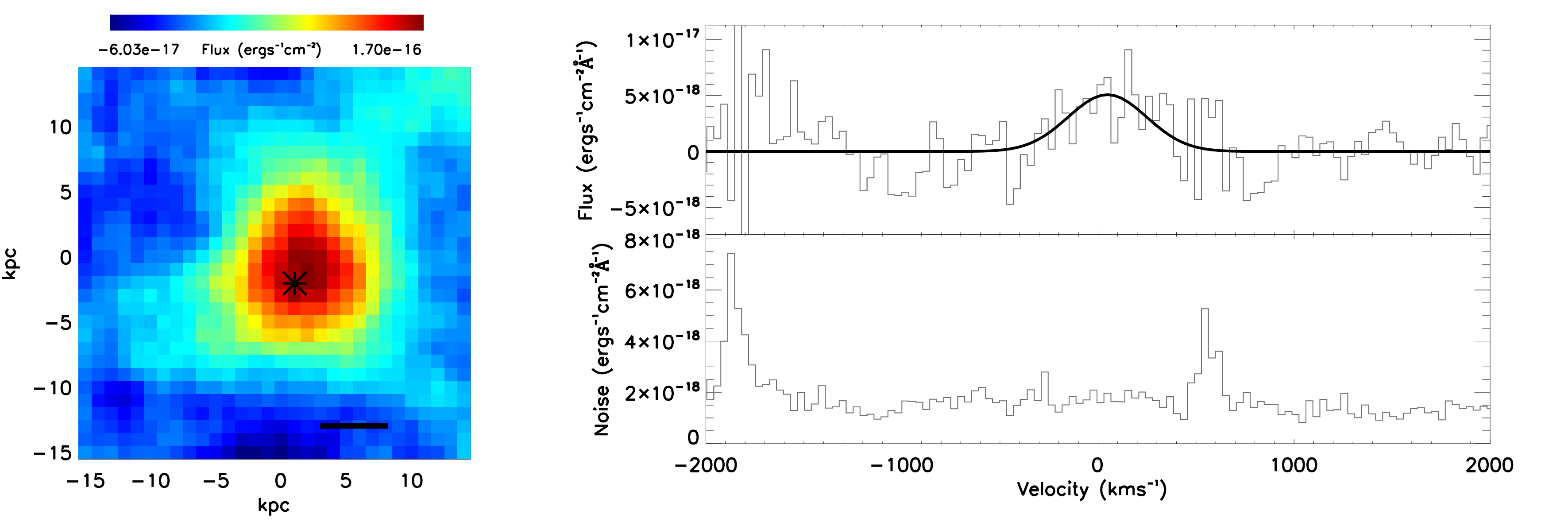}
\caption{As for \protect\ref{fig:narrow_ULASJ1002+0137} for the resolved narrow H$\alpha$ emission in VHSJ2144-0523. A narrow Gaussian profile with a width of $\sigma$=200\,kms$^{-1}$ was found to give a better fit therefore the map is created by integrating over $\pm$300\,kms$^{-1}$ (1.5$\sigma$) and the Gaussian profile overlayed has a width of $\sigma$=200\,kms$^{-1}$. The narrow H$\alpha$ emission is spatially resolved.}
\label{fig:narrow_VHSJ2144-0523}
\end{figure}

\begin{figure}
\includegraphics[width=0.95\textwidth]{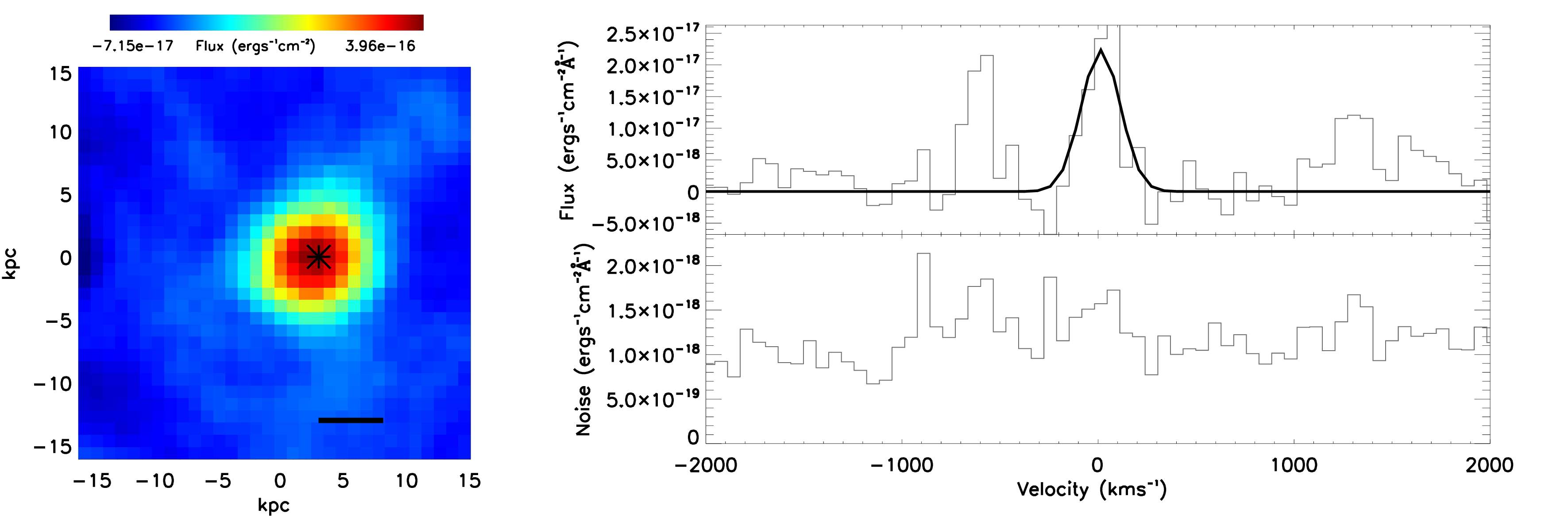}
\caption{As for \protect\ref{fig:narrow_VHSJ1556-0835} for the unresolved narrow H$\alpha$ emission in ULASJ2200+0056. }
\label{fig:narrow_ULASJ2200+0056}
\end{figure}

\begin{figure}
\includegraphics[width=0.95\textwidth]{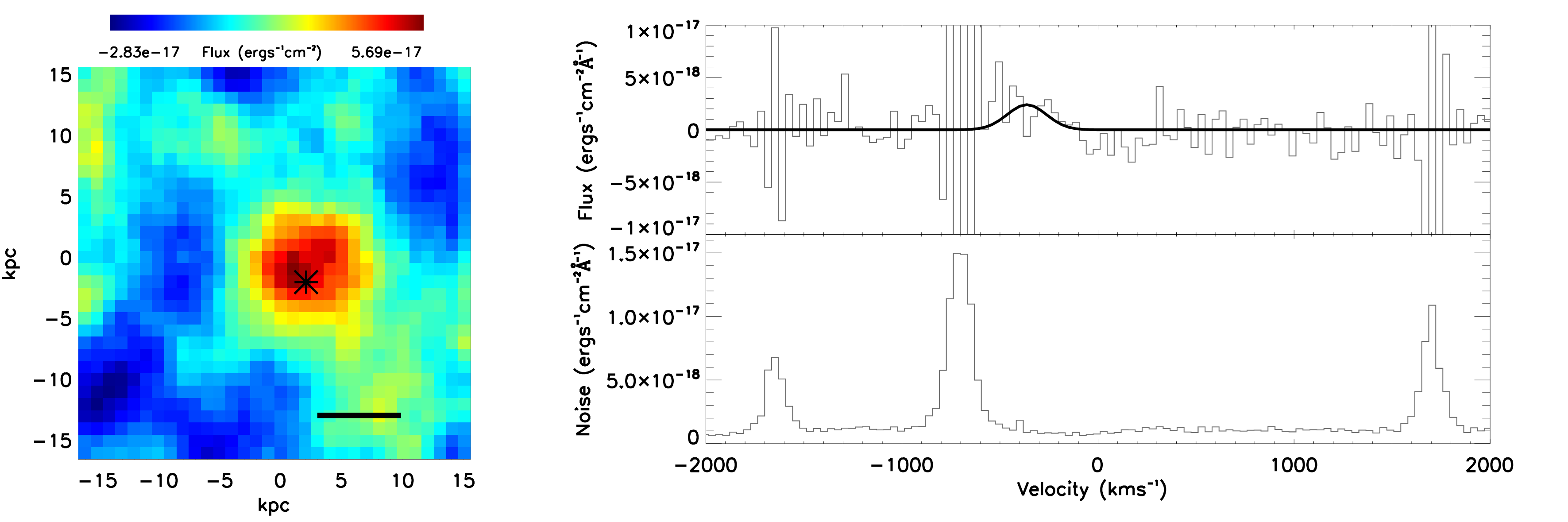}
\caption{As for \protect\ref{fig:narrow_VHSJ1556-0835} for the unresolved narrow H$\alpha$ emission in VHSJ2212-4624. }
\label{fig:narrow_VHSJ2212-4624}
\end{figure}

\begin{figure}
\includegraphics[width=0.95\textwidth]{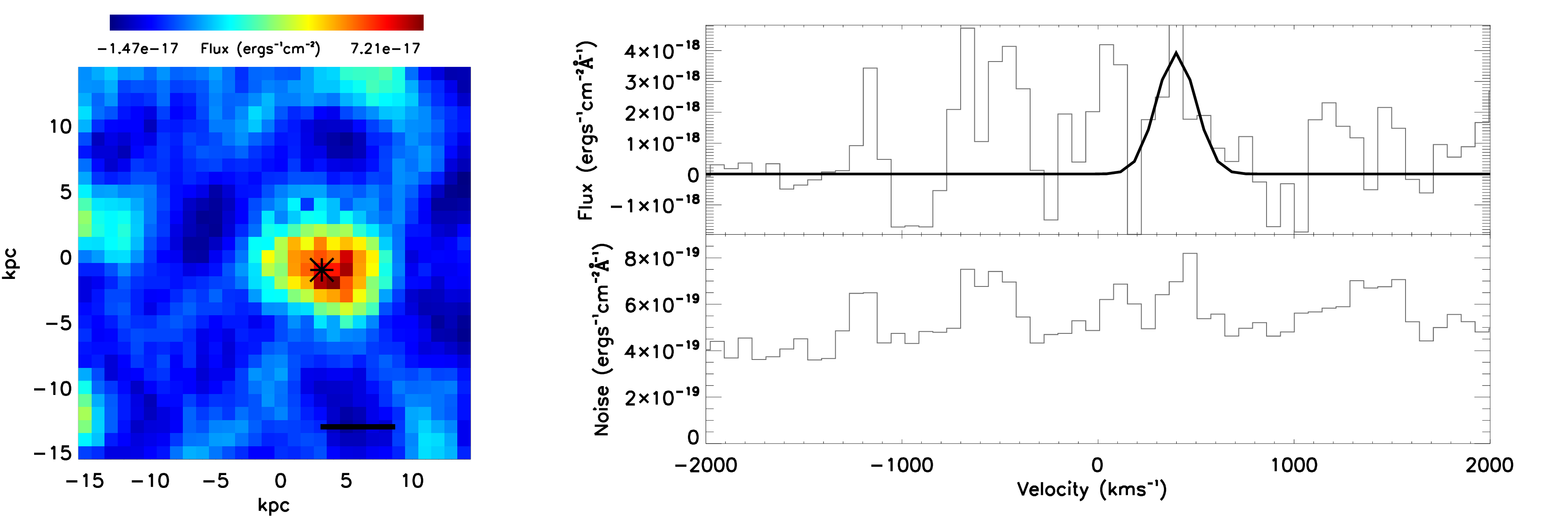}
\caption{As for \protect\ref{fig:narrow_VHSJ1556-0835} for the unresolved narrow H$\alpha$ emission in ULASJ2224-0015. This source is removed from the star formation sample as detailed in Section \protect\ref{sec:crit}.}
\label{fig:narrow_ULASJ2224-0015}
\end{figure}

\begin{figure}
\includegraphics[width=0.95\textwidth]{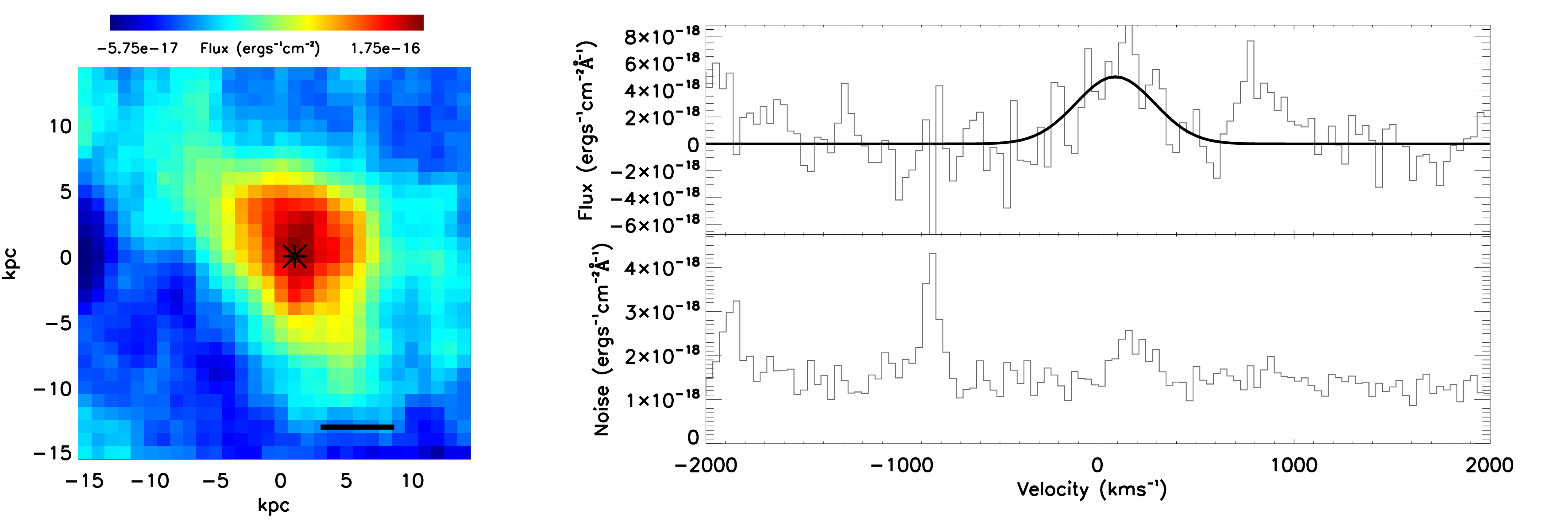}
\caption{As for \protect\ref{fig:narrow_ULASJ1002+0137} for the resolved narrow H$\alpha$ emission in VHSJ2235-5750. A narrow Gaussian profile with a width of $\sigma$=200\,kms$^{-1}$ was found to give a better fit therefore the map is created by integrating over $\pm$300\,kms$^{-1}$ (1.5$\sigma$) and the Gaussian profile overlayed has a width of $\sigma$=200\,kms$^{-1}$. The narrow H$\alpha$ emission is spatially resolved.}
\label{fig:narrow_VHSJ2235-5750}
\end{figure}

\begin{figure}
\includegraphics[width=0.95\textwidth]{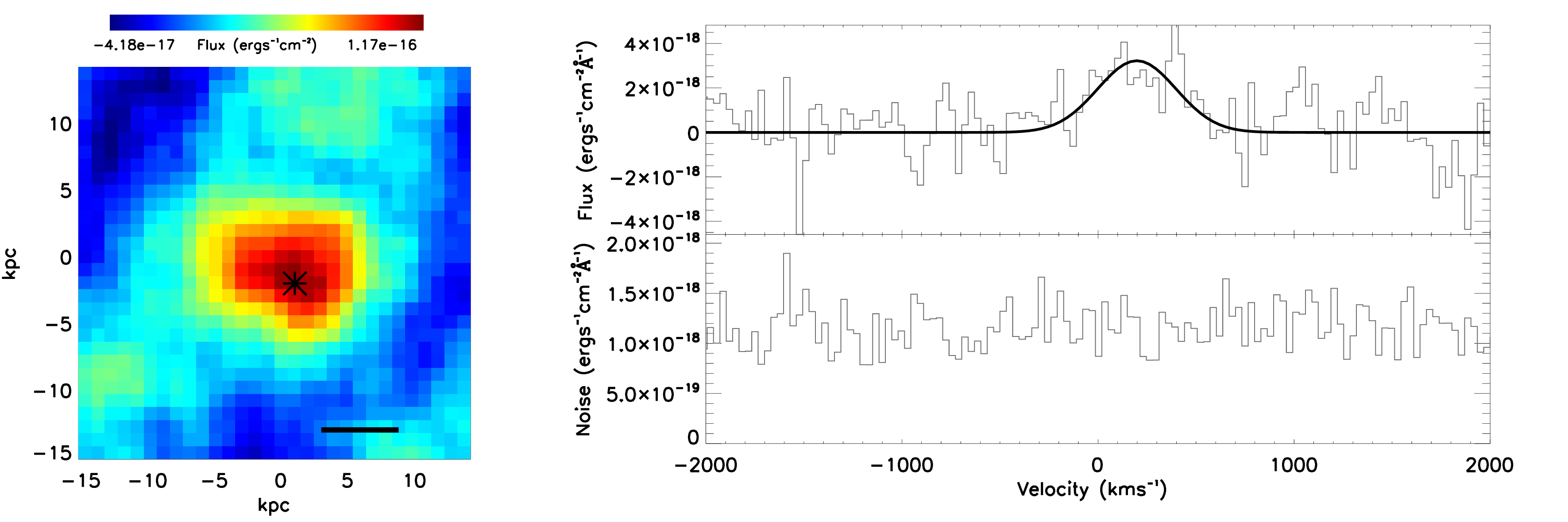}
\caption{As for \protect\ref{fig:narrow_ULASJ1002+0137} for the resolved narrow H$\alpha$ emission in VHSJ2332-5240. A narrow Gaussian profile with a width of $\sigma$=200\,kms$^{-1}$ was found to give a better fit therefore the map is created by integrating over $\pm$300\,kms$^{-1}$ (1.5$\sigma$) and the Gaussian profile overlayed has a width of $\sigma$=200\,kms$^{-1}$. }
\label{fig:narrow_VHSJ2332-5240}
\end{figure}

\begin{figure}
\includegraphics[width=0.95\textwidth]{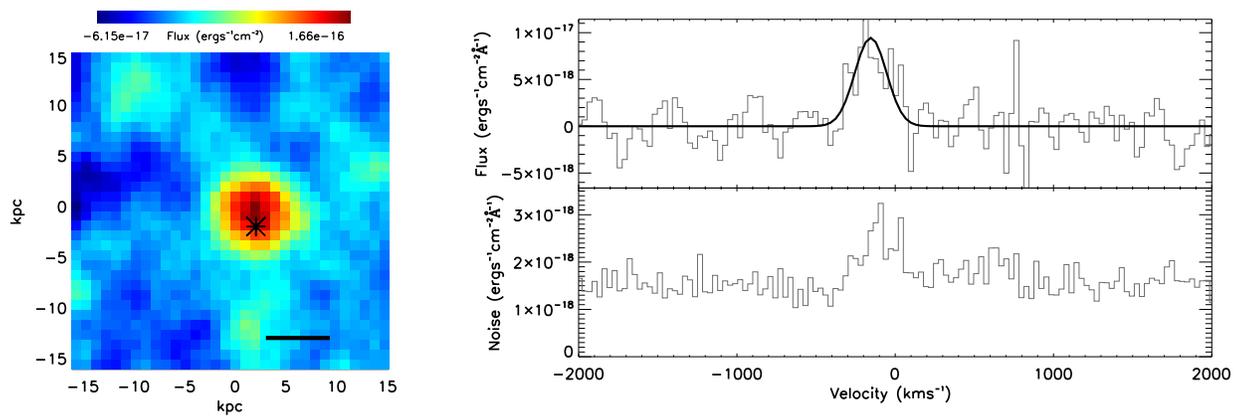}
\caption{As for \protect\ref{fig:narrow_VHSJ1556-0835} for the unresolved narrow H$\alpha$ emission in VHSJ2355-0011. }
\label{fig:narrow_VHSJ2355-0011}
\end{figure}

\end{document}